\documentclass[twocolumn]{aastex631}

\newcommand{\giovani}[1]{\textcolor{black}{#1}}

\usepackage{float}
\usepackage{amsmath}

\usepackage{natbib}
\defcitealias{lazarian1999reconnection}{LV99}
\defcitealias{furth1963finite}{FKR63}

\accepted{for publication in ApJ}

\begin{document}

\title{
Do plasmoids induce fast magnetic reconnection in well-resolved current sheets in 2D MHD simulations?}

\author[0000-0002-7009-9232]{Giovani H. Vicentin}
\affiliation{Departamento de Astronomia, Universidade de São Paulo,
Rua do Matão 1226, 05508-090, São Paulo, Brazil}

\author[0000-0002-0176-9909]{Grzegorz Kowal}
\affiliation{Escola de Artes, Ciências e Humanidades, 
Universidade de São Paulo, Rua Arlindo Bettio 1000, 03828-000, São Paulo, Brazil}

\author[0000-0001-8058-4752]{Elisabete M. de Gouveia Dal Pino}
\affiliation{Departamento de Astronomia, Universidade de São Paulo,
Rua do Matão 1226, 05508-090, São Paulo, Brazil}

\author[0000-0002-7336-6674]{Alex Lazarian}
\affiliation{Department of Astronomy, University of Wisconsin,
475 North Charter Street, Madison, Wisconsin 53706, USA}

%% Note that the \and command from previous versions of AASTeX is now
%% depreciated in this version as it is no longer necessary. AASTeX 
%% automatically takes care of all commas and "and"s between authors names.

%% AASTeX 6.31 has the new \collaboration and \nocollaboration commands to
%% provide the collaboration status of a group of authors. These commands 
%% can be used either before or after the list of corresponding authors. The
%% argument for \collaboration is the collaboration identifier. Authors are
%% encouraged to surround collaboration identifiers with ()s. The 
%% \nocollaboration command takes no argument and exists to indicate that
%% the nearby authors are not part of surrounding collaborations.

%% Mark off the abstract in the ``abstract'' environment. 
\begin{abstract}

We investigate the development of tearing-mode instability using the highest-resolution two-dimensional magnetohydrodynamic simulations of reconnecting current sheets performed on a uniform grid, for Lundquist numbers of $10^3 \le S \le 5 \times 10^5$ , reaching up to $65,536^2$ grid cells. We demonstrate a Sweet--Parker scaling of the reconnection rate $V_{\text{rec}} \sim S^{-1/2}$ up to Lundquist numbers $S \sim 10^4$. For larger values of Lundquist number, between $2\times 10^4\le S \le 2 \times 10^5$, plasmoid formation sets in, leading to a slight enhancement of the reconnection rate, $V_{\text{rec}} \sim S^{-1/3}$, consistent with the prediction from linear tearing mode induced reconnection, indicating that reconnection remains resistivity-dependent and therefore slow. In this range of $S$-values, the plasmoids do not undergo a merger cascade, as they are rapidly advected out of the reconnection layer. Only for $S >  2 \times 10^5$, we observe the nonlinear development of the tearing-mode instability, with plasmoid coalescence and a saturation of the reconnection rate at $V_\text{rec} / V_A \sim 0.01$. At such high $S$, however, the corresponding Reynolds number is large, reaching $\text{Re} > 2000$ even on scales comparable to the current-sheet thickness. We therefore conclude that, in astrophysical systems, it is essential to account for the dominant influence of turbulence and three-dimensional effects in the reconnection process.

\end{abstract}

%% Keywords should appear after the \end{abstract} command. 
%% The AAS Journals now uses Unified Astronomy Thesaurus concepts:
%% https://astrothesaurus.org
%% You will be asked to selected these concepts during the submission process
%% but this old "keyword" functionality is maintained in case authors want
%% to include these concepts in their preprints.
\keywords{Magnetohydrodynamics (MHD) --- Magnetic Reconnection --- Plasmoids --- Methods: Numerical}

%% From the front matter, we move on to the body of the paper.
%% Sections are demarcated by \section and \subsection, respectively.
%% Observe the use of the LaTeX \label
%% command after the \subsection to give a symbolic KEY to the
%% subsection for cross-referencing in a \ref command.
%% You can use LaTeX's \ref and \label commands to keep track of
%% cross-references to sections, equations, tables, and figures.
%% That way, if you change the order of any elements, LaTeX will
%% automatically renumber them.
%%
%% We recommend that authors also use the natbib \citep
%% and \citet commands to identify citations.  The citations are
%% tied to the reference list via symbolic KEYs. The KEY corresponds
%% to the KEY in the \bibitem in the reference list below. 

\section{Introduction} \label{sec:intro}

Magnetic reconnection is a fundamental plasma process in which magnetic field free energy associated with reversals of magnetic field components is transferred to other forms of energy. This process is frequently described in terms of magnetic field topology change, with magnetic lines breaking and reconnecting, allowing the conversion of magnetic energy into kinetic energy, heat, and particle acceleration. This mechanism has long been recognized as playing a central role not only in laboratory plasmas \citep{Taylor1986, Yamada1994, Yamada1997}, but also in a wide range of astrophysical environments, including solar flares \citep{Parker1957, parker1988nanoflares, Masuda1994Nature, PriestForbes2002}, the Earth's magnetosphere \citep{kivelson1995introduction},  proto-stellar disks \citep{dGDP2010, dGDP_2010b}, and relativistic sources such as microquasars,  active galactic nuclei (AGNs), and gamma-ray-bursts (GRBs) \citep[e.g.,][]{dGDP_Lazarian_2005, Giannios2009,dGDP2010, Giannios_2010, zhang-yan2011, nalewajko2011MNRAS.413..333N, Kadowaki_2015, Kadowaki_2021, 
%Ripperda_2020, 
Medina-Torrejón_2021, nishkawa2021,  Medina-Torrejón_2023, dalpino2024}, pulsars \citep[e.g.,][]{Cerutti_2012}, and X-ray binaries \citep[e.g.,][]{Khiali_2015}.

In classical resistive magnetohydrodynamics (MHD), the Sweet--Parker model \citep{Sweet1958, Parker1957} describes the reconnection that occurs in a thin, long current sheet. Although this model provides a self-consistent framework, it predicts a reconnection rate that scales as $V_\text{rec} /V_A \sim S^{-1/2}$, where $S = LV_A/\eta$ is the Lundquist number, $L$ is the typical length scale of the system, $V_A$ is the Alfvén velocity, and $\eta$ is the ohmic resistivity. In highly conducting plasmas ($S \gg 1$), this rate is far too slow to account for the fast reconnection observed in nature, mainly in astrophysical environments, where $S \sim 10^{8-30}$ \citep{Ji_Daughton_2011}. 
From a physical standpoint, the sluggish nature of Sweet--Parker reconnection arises from the strong disparity of the thickness of the matter outflow $\delta$, set by the microscopic (resistive) scale of the Ohmic diffusion and the macroscopic (astrophysical) extent of the surface $L$ over which reconnection takes place. Mass conservation dictates that $V_{rec}\rightarrow 0$ as $\delta/L\rightarrow 0$.

To resolve this discrepancy, 
\cite{petschek1964physics} proposed an alternative model in which the length of the diffusion region ($L$) remains of the order of its thickness ($\delta$). This configuration is achieved through the presence of standing slow-mode shocks, which bend the reconnecting fluxes into an open X-point geometry, thereby enabling fast reconnection rates that depend only weakly on resistivity ($V_\text{rec}/V_A \sim 1 / \ln S$). However, \cite{biskamp1986magnetic} demonstrated numerically that, in resistive MHD with uniform resistivity, the Petschek X-point collapses into elongated current sheets consistent with the Sweet--Parker configuration. As a result, it became widely accepted that fast reconnection requires either non-uniform (anomalous) resistivity or additional physics beyond classical resistive MHD.

Early attempts to sustain Petschek-type reconnection invoked anomalous effects, in particular localized resistivity or two-fluid contributions such as the Hall term \citep[e.g.][]{Ugai_1992, Shay_1999, Birn_2001}. While Hall reconnection remains a key framework in collisionless space and laboratory plasmas, numerical investigations over the following decade indicated that such mechanisms alone could not provide a universal explanation for fast reconnection in large-scale astrophysical environments, where kinetic scales are negligible compared with macroscopic system sizes. As a result, the focus of the astrophysical community shifted toward the plasmoid--mediated tearing instability as the mechanism capable of enabling fast reconnection in two-dimensional resistive MHD \citep{loureiro2007instability, bhattacharjee2009fast, Uzdensky_2010}.\footnote{We do not address here the theory of 3D turbulent reconnection, where the current sheet thickness $\delta$ is set by magnetic-field line wandering in the turbulent flow and can be comparable with $L$ \citep{lazarian1999reconnection}. The predictions of this theory have been successfully confirmed numerically \citep{kowal2009numerical, kowal2012visc, Vicentin_2025} and applied to explain a broad range of space-physics and astrophysical observations \citep[see, e.g.][for a review]{Lazarian2020review}.}

Attempts to increase $\delta$ by appealing to instabilities have been explored in the literature. Resistive instabilities in a magnetized fluid can trigger a long‐wavelength tearing-mode within the current sheet, driving magnetic reconnection and spawning a chain of narrow magnetic islands in a two‐dimensional flow, as first demonstrated by \cite{furth1963finite}.  Building on the notion that a continuous sheet fragments into discrete islands, \cite{heyvaerts1984coronal} argued that such tearing modes might underpin the heating of the solar corona, while \cite{lee1985theory, lee1986multiple}  extended the idea to Earth’s magnetopause by modeling reconnection along multiple X-lines. Extensive work by S. Syrovatskii and collaborators proposed that tearing instability is a generic feature of astrophysical reconnection \citep[see][and references therein]{Syrovatskii_1981}.

\citet{lazarian1999reconnection}, considering steady-state reconnection on the scale of the reconnection sheet $L$, demonstrated that the linear regime of tearing instability increases $\delta$ (see their Appendix C), yielding a reconnection rate that scales as $S^{-3/10}$. Although this represents an improvement over the Sweet--Parker rate, it remains far too slow to account for astrophysical reconnection.

The interest in the role of tearing instability for reconnection has been renewed more recently due to the increase in available computational power. The primary object of such studies was 2D reconnection, where higher numerical resolution is available. In this paper, we follow the trend, even though the physics of reconnection can be different in 2D and 3D.  

2D numerical simulations revealed the transition from the Sweet--Parker reconnection to a regime where
tearing-mode instability becomes significant. In particular, it was observed that when the current sheet is sufficiently thin, the tearing instability develops. A new effect,
absent in earlier theoretical models,
was observed in the numerical simulations,  namely, the formation of chains of magnetic islands within the 2D current sheet \citep{loureiro2007instability}. 
These islands, commonly referred to as plasmoids, were observed to interact, merge, and grow into so-called ``monster'' plasmoids \citep{Uzdensky_2010, Loureiro2012}, which are subsequently advected away by the 2D outflow.
This 2D cascade of growing plasmoids was assumed to enable what earlier tearing-mode theoretical studies could not accomplish, namely, the combination of tearing reconnection at the smallest resistive scales with the efficient outflow of matter from the reconnection region on the scale of the ``monster'' plasmoids. The size of these structures determines the thickness of the outflow and may not depend on resistivity, potentially enabling $S$-independent reconnection.

The subsequent studies with dedicated 2D MHD simulations \citep{bhattacharjee2009fast, huang2010scaling, Loureiro2012} 
reported that for high Lundquist numbers, $S \gtrsim 10^4$, plasmoids can induce reconnection independent of Ohmic resistivity $\eta$, at a universal rate of $V_\text{rec} \sim 0.01 \, V_A$, which is the expected Sweet--Parker rate for the critical Lundquist number of $S_c = 10^4$.

Applying proper boundary conditions is nontrivial in reconnection simulations. To overcome this difficulty, \cite{bhattacharjee2009fast}  
adopted two coalescing magnetic flux tubes \citep{uzdensky2000two} for the reconnection setup. The authors reported that in their simulations, plasmoids are formed when Lundquist numbers reach a critical value of $S \gtrsim 3 \times 10^4 = S_c$. In \cite{huang2010scaling}, the authors repeated the 2D simulations with higher resolution and different noise amplitudes ($\epsilon$), and observed that the critical Lundquist number can reach up to $S_c=10^5$ for $\epsilon = 10^{-5}$. Expanding this analysis, \cite{Huang_2017} calculated the critical Lundquist number as a function of the initial noise, and 
found that $S_c = 10^6$ for $\epsilon = 10^{-30}$ (see their Fig. 10).

More recently, \cite{Morillo2025} conducted high-resolution 2D MHD simulations of the \cite{Orszag_Tang_1979} vortex with a uniform-grid to investigate magnetic reconnection processes and the development of tearing-mode instability. They found that when the current layer is well-resolved, specifically when the ratio between the thickness of the current sheet ($\delta$) and the simulation's cell size ($h$) satisfies $\delta /h \ge 10$, plasmoid formation is suppressed. 
\cite{Morillo2025} observed that, under these well-resolved conditions, the reconnection rate adheres to the Sweet--Parker model, indicating a slow Lundquist number-dependent reconnection. Their simulations demonstrated that even at a Lundquist number as high as $5 \times 10^5$, achieved with a resolution of $32,768^2$ grid points, no plasmoid instability developed.

The work by \cite{Morillo2025} clearly demonstrated the importance of resolving the details of the current sheet for studies of tearing reconnection.
However, the tearing-mode instability is a genuine physical process observed both in nature and in laboratory plasma experiments \citep[e.g.,][]{Jara_Almonte_2016}. This makes it essential to explore the development of the instability, keeping the resolution of the current sheet high.

To understand the nature of high Lundquist number 2D reconnection, 
we perform simulations that satisfy the \cite{Morillo2025} criterion, of $\delta /h>10$, but we add
a small level of noise to 
ensure that the instability develops, avoiding the possibility that it is suppressed by an insufficient initial perturbation. In our simulations, we use the criterion $\delta/h>10$ as an empirical guideline and minimum threshold, but we supplement it with a theoretically motivated inner-layer criterion derived from linear tearing theory, and we estimate the numerical error of the simulations using the magnetic-energy balance diagnostic.

In Section \ref{Sec:Linear_inst_TM}, we present a review of the linear theory of the tearing-mode instability, which provides the theoretical foundation for our study.
Section \ref{sec:numerics} describes the numerical setup of the simulations and the computational methods implemented. In Section \ref{sec:recrate}, we present the procedure used to measure the reconnection rate in our 2D MHD simulations. Section \ref{sec:results} contains the main results, with particular emphasis on convergence analysis between different numerical resolutions, as well as the introduction of a new method to quantify numerical errors directly from the simulations. In Section \ref{sec:discussion}, we summarize and discuss the implications of our findings and in Section \ref{sec:conclusions}, we present the main conclusions of the paper.

\section{Tearing-Mode Instability and the Critical Lundquist Number}
\label{Sec:Linear_inst_TM}

\subsection{Linear Theory of the Tearing Instability}
\label{ssec:Linear_Tearing}

The tearing instability of thin current sheets has been the subject of detailed analysis since the seminal works of \citet{furth1963finite} and \citet{Coppi1976}. It arises when a magnetic configuration with antiparallel field lines develops a thin resistive layer that permits magnetic reconnection, causing exponential growth of perturbations on top of the equilibrium sheet. The character of the instability depends on how magnetic flux is redistributed across the inner resistive layer, and two asymptotic regimes can be distinguished depending on the parameter $ka S_a^{1/4}$, where $a$ is the sheet half–thickness, $k$ the perturbation wavenumber along the sheet, and $S_a = V_A a/\eta$ the Lundquist number defined with respect to $a$. Here, $V_A$ is the upstream Alfvén speed and $\eta$ the magnetic diffusivity.

When $ka S_a^{1/4}\gg 1$ (known as \emph{constant-$\psi$} regime), the magnetic flux function can be regarded as constant across the inner resistive layer. In this case, the growth rate of the instability scales as
\begin{equation}
\gamma_{\rm FKR} \sim \frac{V_A}{a} \, S_a^{-3/5} (ka)^{-2/5},
\label{eq:FKR}
\end{equation}
the classical result of \citet[][hereafter \citetalias{furth1963finite}]{furth1963finite}. Physically, this regime corresponds to relatively short-wavelength modes, where the outer region responds rigidly. As a result, flux is reconnected slowly and the instability growth is strongly damped by resistivity.

When instead, $ka S_a^{1/4}\ll 1$, the flux function is no longer constant across the inner layer. In this regime (known as \emph{non-constant-$\psi$} regime), perturbations can redistribute magnetic flux more effectively, leading to a faster growth rate
\begin{equation}
\gamma_{\rm Coppi} \sim \frac{V_A}{a} \, S_a^{-1/3} (ka)^{2/3},
\end{equation}
as obtained by \citet{Coppi1976}. This regime applies at longer wavelengths, where the sheet is more easily destabilized because magnetic tension is weaker over larger scales.  

The transition between these two regimes is set by the condition $\Delta' \delta_{\rm in} \sim 1$, where $\Delta'$ is the tearing stability parameter from outer-region matching, and $\delta_{\rm in}$ is the resistive inner layer width. For a Harris-type sheet, this condition yields the crossover scale
\begin{equation}
k_\ast a \sim S_a^{-1/4}.
\end{equation}
At this wavenumber the instability achieves its maximum linear growth rate,
\begin{equation}
\gamma_{\max} \approx C_\gamma \frac{V_A}{a}\,S_a^{-1/2}, 
\label{eq:gamma_max}
\end{equation}
with corresponding wavenumber
\begin{equation}
k_{\max} \approx \frac{C_k}{a\,S_a^{1/4}}.
\label{eq:kmax}
\end{equation}
The coefficients $C_\gamma$ and $C_k$ are of order unity but depend somewhat on the equilibrium profile and boundary conditions. Our numerical analysis for $Pr_{\rm m}=1$ gives $C_\gamma \approx 0.5$ and $C_k \approx 1.05$ (see Appendix~\ref{sec:appendix_growth}), which are consistent with earlier work \citep[e.g.][]{loureiro2007instability, Tenerani_2015}.  

Viscous and compressible effects further enrich this picture. Early work by \citet{porcelli1987viscous} showed that viscosity can significantly alter growth rates and inner-layer scalings when the magnetic Prandtl number $Pr_{\rm m} = \nu/\eta$ is large. Subsequent studies, such as \citet{Tenerani_2015}, developed a systematic treatment of the visco-resistive tearing instability in Harris-type current sheets. For the case of $Pr_m=1$ considered here, however, viscosity modifies only the numerical prefactors, 
leaving the classical scaling laws essentially unchanged.

From a physical perspective, the key point is that thin current sheets are generically unstable. For a fixed sheet half-thickness $a$, the maximum tearing growth rate scales as $\gamma_{\max} \propto S_a^{-1/2}$, i.e. it decreases with the local Lundquist number $S_a$. However, in a Sweet--Parker sheet the thickness itself shrinks with the global Lundquist number as $a \sim L S^{-1/2}$, so that the effective local parameter scales as $S_a \sim S^{1/2}$. Substituting this relation into the expression for $\gamma_{\max}$ yields the well-known result $\gamma_{\max} \tau_A \propto S^{1/4}$ \citep{loureiro2007instability}, meaning that in practice the instability becomes stronger as $S$ increases. Thus, beyond a critical global Lundquist number, any sufficiently long sheet will inevitably fragment into multiple plasmoids, and the linear tearing modes become the seeds for nonlinear reconnection.

In our numerical simulations we employ two complementary strategies to excite these tearing modes. First, we apply a small-amplitude random noise perturbation to the velocity field. This approach excites a broad spectrum of wavenumbers, allowing the system to select and amplify the fastest-growing tearing harmonics predicted by linear theory. Second, we impose multimode perturbations directly in Fourier space, deliberately choosing wavenumbers around $k_{\max}$ from Eq.~(\ref{eq:kmax}). In this case, the seeded modes are those expected to grow most rapidly according to theory, ensuring that the instability develops efficiently and shortening the linear stage. These perturbation strategies therefore provide both a generic and a targeted way to trigger tearing, directly reflecting the analytical estimates of unstable modes.

\subsection{Reconnection Rate from the Maximum--Growth Tearing Mode}
\label{ssec:Tearing_Scaling}

\citet{lazarian1999reconnection} estimated the global reconnection rate by equating the tearing growth rate to the shear outflow rate. In their Appendix~C, they employed the
\citetalias{furth1963finite} growth rate (Eq. \ref{eq:FKR}), obtaining
\begin{equation}
  V_{\rm rec} \propto V_A S^{-3/10},
  \label{eq: vrec_LV}
\end{equation}
with $S \equiv V_A L/\eta$. If instead the fastest growing tearing mode at the FKR--Coppi crossover is considered, then the maximum growth rate from Eq.~(\ref{eq:gamma_max}) applies. Importantly, the growth rate $\gamma_{\max}$ increases as the sheet half-thickness $a$ decreases. However, as argued by \cite{lazarian1999reconnection}, a steady-state reconnection layer requires that mass outflow from the current sheet proceed at a speed limited by $V_A$. 

This constraint implies that the inflow speed (and hence the reconnection rate) cannot grow arbitrarily as $a$ shrinks. Rather, mass conservation requires the reconnection rate to scale with the ratio $a/\lambda_\parallel$, where $\lambda_\parallel$ denotes the parallel scale of plasma exhaust along the sheet. In other words, reconnection becomes faster with larger $a$, while the resistive reduction of $a$ is balanced by the need to sustain an Alfvénic outflow. This situation is analogous to the Sweet--Parker case, where reducing $a$ increases resistive efficiency but is constrained by mass conservation through the global outflow bottleneck.

In terms of the outflow rate $\gamma$, the steady-state condition can be written as
\begin{equation}
  \gamma \sim \frac{V_A}{\lambda_\parallel}, \qquad V_{\rm rec} \sim \gamma\,a \sim V_A\frac{a}{\lambda_\parallel}.
\end{equation}
While the maximum linear tearing mode has a characteristic wavelength $\sim k_{\max}^{-1}$ that governs the {\em spacing of plasmoids}, the global reconnection rate is set by the overall exhaust, for which the appropriate parallel scale is the system size, $\lambda_\parallel \sim L$. Substituting $\gamma_{\max}$ (Eq. \ref{eq:gamma_max}) with this identification yields
\begin{equation}
  \frac{V_A}{\lambda_\parallel} \;\sim\; \frac{V_A}{a}\,S_a^{-1/2} = \frac{V_A}{a}\left(\frac{\eta}{V_A a}\right)^{1/2},
\end{equation}
which provides $a$:
\begin{equation}
  a \;\sim\; \left(\frac{\eta\,\lambda_\parallel^2}{V_A}\right)^{1/3}.
\end{equation}
Hence, the global {\it steady-state} reconnection rate becomes
\begin{equation}
V_{\rm rec} \sim V_A\left(\frac{\eta}{V_A L}\right)^{1/3} \sim V_A S^{-1/3}.
\label{eq:Vrec_max}
\end{equation}

Thus, evaluating the reconnection rate at the maximum tearing growth mode, but using the global parallel scale for the outflow constraint, produces the tearing-limited scaling $V_{\rm rec}\propto V_A S^{-1/3}$. The internal wavenumber $k_{\max}$ sets the number and distribution of plasmoids, whereas the global throughput remains limited by the system-length exhaust.

The derivation presented here and in \citet{lazarian1999reconnection} differs from other analyses of the tearing reconnection rate \citep[see e.g.][]{Loureiro2009}, as it incorporates a self-consistent treatment of both the parallel and perpendicular scales of tearing—an essential requirement for achieving steady-state reconnection. In particular, \citet{lazarian1999reconnection} and the derivation above require that the perpendicular scale of tearing correspond to the parallel extent of the current sheet.

Eq.~(\ref{eq:Vrec_max}) resembles the Sweet--Parker prediction, as it retains an explicit dependence on the Lundquist number $S$, though with a slightly modified exponent. As a result, linear tearing-limited reconnection proceeds somewhat faster than Sweet--Parker, but still remains slow for astrophysically relevant $S$. Using only the \citetalias{furth1963finite} branch within the same shear--decorrelation framework gives $V_{\rm rec}\propto V_A S^{-3/10}$ \citep{lazarian1999reconnection}, whereas allowing the spectrum to reach the FKR--Coppi crossover (maximum growth) yields the slightly steeper $V_{\rm rec}\propto V_A S^{-1/3}$. The differences between these two scalings are of secondary practical importance. In both cases the physical interpretation is the same---reconnection is limited by shear stabilization and constrained by mass conservation, with $V_{\rm rec}\sim V_A a/\lambda_k$; the only distinction is whether the growth rate is evaluated on the FKR branch or at the FKR--Coppi crossover. The transition from the Sweet--Parker scaling to the faster tearing-limited scaling of Eq.~(\ref{eq:Vrec_max}) is expected as soon as the current sheet becomes unstable to tearing.

\subsection{Critical Lundquist Number for the Transition to Plasmoid-Mediated Reconnection}
\label{ssec:Critical_Sc}

A Sweet--Parker (SP) current sheet of length $L$ has half-thickness $a_{\rm SP} \sim L S^{-1/2}$, corresponding to $S_a\sim S^{1/2}$. In such a sheet, the fastest tearing mode grows at
\begin{equation}
k_{\max}L \approx C_k S^{3/8}, \quad \gamma_{\max}\tau_A \approx C_\gamma S^{1/4}, 
\quad \tau_A=L/V_A.
\label{eq:SPmax}
\end{equation}

The plasmoid instability requires that unstable modes both fit within the sheet and grow sufficiently before being advected out. The first condition, $k_{\max}L \gtrsim 1$, is already met at very low $S$ and is therefore not a limiting constraint. The second condition, $\gamma_{\max}\tau_A \gtrsim N$, expresses that the perturbation amplitude must grow by $N$ e--foldings\footnote{The ``e--folding'' time is the interval over which an exponentially growing or decaying quantity changes by a factor of $e\simeq 2.718$. The terminology is standard in astrophysics and fluid dynamics; see, e.g., \citet{Lipps1963} or \citet{Goldsmith1970}.} during one Alfvén crossing time. Writing the perturbation as $A(t) = A_0 e^{\gamma t}$, a single e-fold corresponds to amplification by a factor of $e$, while $N=5$ corresponds to a factor of $\sim 150$, and $N=10$ to $\sim 2.2\times 10^4$. In practice, several e--folds are required for perturbations to rise above background noise and drive nonlinear disruption. This condition leads to the threshold
\begin{equation}
S_c \approx \left(\frac{N}{C_\gamma}\right)^4.
\label{eq:sc_crit}
\end{equation}
For $C_\gamma\approx 0.5$, this simplifies to $S_c \approx (2N)^4$, so that $N=5$--10 yields $S_c \sim 10^4$--$10^5$.

It is also important to recall the long-wavelength ordering for tearing, $ka < 1$, which is a prerequisite for the applicability of the linear theory. For the fastest tearing mode in a SP current sheet we have
\begin{equation}
k_{\max} a = (k_{\max}L)\frac{a}{L} \approx C_k S^{3/8} S^{-1/2} = C_k S^{-1/8}.
\end{equation}
Thus the criterion $k_{\max} a < 1$ is automatically satisfied for $S \gg 1$, since $k_{\max}a$ decreases as $S^{-1/8}$. The corresponding bound is
\begin{equation}
S \gtrsim S_{ka} \equiv C_k^{\,8},
\end{equation}
which with $C_k \approx 1.05$ gives $S_{ka} \approx 1.5$. This value is negligible compared to the thresholds implied by the growth condition. More generally, the interval of unstable wavenumbers satisfying $L^{-1} \lesssim k \lesssim a^{-1}$ always exist for SP current sheets with $a<L$. Therefore, in practice, the onset of plasmoid formation is not limited by the $ka<1$ ordering, but by the requirement of sufficient exponential amplification.

Thus, the analytic estimate indicates that plasmoid--mediated reconnection should onset at $S_c \sim 10^4$--$10^5$, corresponding to several e--folds of growth within one global Alfvén time. In the next sections we will examine how this theoretical threshold compares with our numerical simulations.

\section{Numerical Methodology} \label{sec:numerics}
\subsection{The code}

We use the high-order shock-capturing Godunov-type code AMUN\footnote{The code is freely available at \url{https://bitbucket.org/amunteam/amun-code/}.} \citep{kowal2009numerical, kowal2012visc} to solve the isothermal visco-resistive 2D MHD equations:

\begin{equation}
    \frac{\partial \rho}{\partial t} + \nabla \cdot (\rho \mathbf{v}) = 0,
\end{equation}

\begin{equation}
    \frac{\partial (\rho \mathbf{v})}{\partial t} + \nabla \cdot \left[ \rho \mathbf{v}\mathbf{v} + \left( p + \frac{B^2}{8\pi} \right) \mathbf{I} - \frac{1}{4\pi} \mathbf{B} \mathbf{B} \right] = \nabla \cdot \mathbf{\tau} + \mathbf{f},
\end{equation}

\begin{equation}
    \frac{\partial \mathbf{B}}{\partial t} + \nabla \times \mathbf{E} = 0,
\end{equation}

\noindent
where $\rho$ and $\mathbf{v}$ are the plasma density and velocity, respectively, $\mathbf{B}$ is the magnetic field, $\mathbf{E} = - \mathbf{v} \times \mathbf{B} + \eta \mathbf{j}$ is the electric field, $\mathbf{j} = \nabla \times \mathbf{B}$ is the current density, $p = \rho c_s^2$ is the thermal pressure, $c_s$ is the isothermal sound speed, $\eta$ is the resistivity coefficient, $\mathbf{\tau} = \nu \rho \left[ \nabla \mathbf{v} + \nabla^T \mathbf{v} - \frac{2}{3} \nabla \cdot \mathbf{v} \right]$ is the viscous stress tensor, $\nu$ is the kinematic viscosity, and $\mathbf{f}$ represents the forcing term. 

To numerically solve the 2D MHD set of equations, we used a \emph{kinetic energy preserving} and \emph{entropy stable} (KEPES) Riemann solver \citep{Derigs_etal:2018} with a $7^\mathrm{th}$-order Monotonicity-Preserving (MP7) reconstruction method \citep{mp5} to reconstruct the Riemann states, and a $3^\mathrm{rd}$-order 4-step Embedded Strong Stability Preserving Runge-Kutta (SSPRK) method for time advance, where the time step is controlled by both the Courant–Friedrichs–Lewy (CFL) condition and the integration error \citep[see, e.g.,][]{SSPRK324}.

The code uses dimensionless equations in such a way that the strength of the magnetic field is expressed in terms of the Alfvén velocity, which is defined by the antiparallel component of the magnetic field (reconnecting field) and the unperturbed density $\rho_0 = 1$. All other velocities are expressed as units of the Alfvén speed, the length of the box in the $x-$direction ($L_x$) defines the unit of distance, and time is measured in units of the Alfvén time, defined as $t_A \equiv L_x/V_A$.

\subsection{Numerical setup}

\begin{figure}[ht]
    \centering
    \includegraphics[width = 0.49 \textwidth]{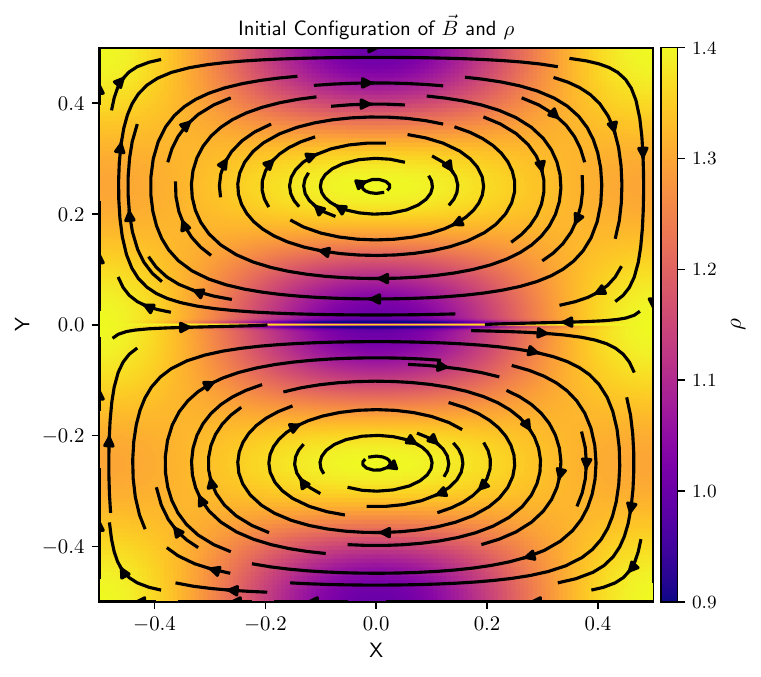}
    \caption{Initial configuration of magnetic field and density for the simulation with $S=10^5$ and $\beta = 2.0$. Black arrows represent the in-plane component of the magnetic field, and the colormap is the density profile. The out-of-plane component of the magnetic field $(B_z)$ is set to be constant.
    }
    \label{fig:initial-config}
\end{figure}

The configuration of the magnetic field in the reconnection region is similar to the one employed by \cite{bhattacharjee2009fast, huang2010scaling}, where the attraction between two coalescing magnetic flux tubes is the driver of magnetic reconnection. The simulation domain is a 2D square with dimensions $L_x = L_y = L = 1$, and the effective resolution for the simulations varies between $512^2$ and $65,536^2$ grid points. 
The reconnecting magnetic field is  along the $x-$direction (see Fig. \ref{fig:initial-config}). 
The boundary conditions in our models are perfectly conducting, free slipping boundaries along $x$ and $y$ directions.

The initial magnetic field is given by $\mathbf{B} = \hat{z} \times \nabla \psi + B_z \, \hat{z}$, where 

\begin{equation}
    \psi = \frac{1}{2 \pi} \tanh \left(\frac{y}{\delta} \right) \cos (\pi x) \sin (2\pi y), \label{eq:psi_HB}
\end{equation}

\noindent
and $B_z$ is the guide field. We adopted, for our initial configuration, a constant guide field with amplitude $B_z = 0.5$, while the density is nonuniform to maintain the pressure balance (see Fig. \ref{fig:initial-config}).  In all simulations performed, the plasma$-\beta$, the ratio between thermal and magnetic pressure, is fixed at $\beta = p_\text{th}/p_\text{mag} = 2 \rho c_s^2/B^2 = 2.$

The initial density profile (shown as the colormap in Fig. \ref{fig:initial-config}) is constructed from the pressure balance $p_{\rm tot} = p_{\rm mag} + p_{\rm th}$, assuming $p_{\rm tot}$ constant, with $p_{\rm th} = \rho c_s^2$. The constant $p_{\rm tot} = p_{\rm mag, max} + p_0$ is chosen so that the initial state is in total-pressure equilibrium across the entire domain and the density remains positive everywhere, with $p_{\rm mag, max} = \frac12 (B_0^2 + B_z^2) $, $B_0 = 1$, and $p_0 = \beta p_{\rm mag,max}$.

The initial thickness of the current sheet is set to $\delta = LS^{-1/2}$, following the Sweet--Parker (SP) scaling. At lower Lundquist numbers, where reconnection is expected to follow the SP scaling, this prescription provides a natural and consistent starting point. More importantly, because our objective is to identify the range of $S$ in which the behavior departs from the Sweet--Parker regime, initializing the system with an SP-scaled sheet ensures that any deviation observed in the reconnection dynamics is attributable to the physics of tearing and not to variations in the initial geometry.

As stressed, in our simulations we employ a uniform grid, as in the case of \cite{Morillo2025}, so the resolution across the current sheet and all secondary current layers does not degrade away from the centre, and we introduce an explicit convergence analysis based on the magnetic-energy balance and on resolving the inner resistive layer of the tearing-mode instability.

In previous works, e.g., \cite{bhattacharjee2009fast, huang2010scaling, Huang_2017}, the authors mention the minimum grid size, but it is unclear how the mesh refinement works, and therefore, the information about how many cells are resolving the current sheet is missing. On the other hand, in order to achieve higher Lundquist numbers, of the order of $S = 10^6 - 10^9$, the initial laminar current layer is so thin that using a uniform grid is extremely computationally expensive, and adopting a mesh refinement is necessary.

\subsection{Initial perturbation} \label{subsec:turbulence}

We have tested two different methods to drive the initial perturbation in the system. First, we adopted an initial Gaussian noise in the velocity field, with amplitudes of $\delta v = \{ 1, 10, 100 \} \times 10^{-3} \, V_A$. 

We also drove small-scale perturbation for a short initial period into the system. In this case, we employ a technique described by \cite{alvelius1999random} \citep[see also][for a detailed discussion of this method]{kowal2009numerical, Kulpa-Dybel-2010}. This forcing is applied in the spectral space, concentrated around a wave vector $k_{\text{inj}}$ that corresponds to the injection scale $l_{\text{inj}} \sim k_\text{inj}^{-1}$. Within a shell extending from $ k_\text{inj} - \Delta  k_\text{inj}$ to $ k_\text{inj} + \Delta  k_\text{inj}$, we disturb $N$ discrete Fourier components of velocity using a Gaussian profile characterized by a half-width $k_c$ and a peak amplitude $v_f$ at the injection scale. The amplitude of the perturbation is given by the injection power $P_\text{inj}$.

We adopted different combinations of $k_\text{inj}$ and $P_\text{inj}$, but in all models tested with external forcing, the perturbation is injected into the system at the beginning of the simulation ($t=0$) up to $t=0.1 \, t_A$. After that, the simulations evolve without external perturbation. Table \ref{tab:list_models} lists the initial conditions of the models simulated in this work.

\begin{table}[t]
\centering
\begin{tabular}{cccccccc}
$S (\times 10^{3})$ & $h^{-1}$  & Perturb. & $\delta v$ & $k$ &  \giovani{$t_\text{max}$} \\ \hline \hline
1 & 512 & RN & $10^{-2}$ & -  & {3.0} \\
2 & 512 & RN & $10^{-2}$ & - & {3.0}   \\
5 & 512 & RN & $10^{-2}$ & - & \giovani{5.0}  \\
5 & 1024 & RN & $10^{-2}$ & - & \giovani{5.0}  \\
10 & 512 & RN & $10^{-2}$ & - & \giovani{5.0}  \\
10 & 1024 & RN & $10^{-2}$ & - & \giovani{5.0}  \\
10 & 2048 & RN & $10^{-2}$ & - & \giovani{5.0} \\ \hline
20 & 1024 & RN & $10^{-2}$ & - & \giovani{5.0} \\ 
20 & 2048 & RN & $10^{-2}$ & - & \giovani{5.0} \\ 
33 & 1024 & RN & $10^{-2}$ & - & \giovani{5.0} \\ 
33 & 2048 & RN & $10^{-2}$ & - & \giovani{5.0} \\ \hline
50 & 512 & RN & $10^{-2}$ & - & \giovani{5.0} \\
50 & 1024 & RN & $10^{-2}$ & - & \giovani{5.0} \\
50 & 2048 & RN & $10^{-2}$ & - & \giovani{5.0}  \\
50 & 4096 & RN & $10^{-3}$ & - & \giovani{5.0}  \\
50 & 4096 & RN & $10^{-2}$ & - & \giovani{5.0}  \\
50 & 4096 & RN & $10^{-1}$ & - & \giovani{5.0}  \\
50 & 8192 & RN & $10^{-2}$ & - & \giovani{5.0} \\ \hline
50 & 8192 & MM & - & 60 & \giovani{5.0}  \\
50 & 8192 & MM & - & 128 & \giovani{5.0}  \\
50 & 8192 & MM & - & 256 & \giovani{5.0}  \\
50 & 8192 & MM & - & 1024 & \giovani{5.0}  \\ \hline
100 & 8192 & RN &  $10^{-2}$ & - & \giovani{5.0}  \\
100 & 16384 & RN &  $10^{-2}$ & - & \giovani{10.0}  \\
100 & 16384 & MM & - & 80 & \giovani{5.0}  \\ \hline
200 & 32768 & RN & $10^{-2}$ & - & \giovani{5.0}  \\
250 & 32768 & RN & $10^{-2}$ & - & \giovani{3.0}  \\
333 & 32768 & RN & $10^{-2}$ & - & \giovani{1.2}  \\
333 & 65536 & RN & $10^{-2}$ & - & \giovani{1.2} \\
500 & 65536 & RN & $10^{-2}$ & - & \giovani{1.2}
\end{tabular}

    \caption{List of initial parameters of the simulated models. In this table, $S = LV_A/\eta$ is the Lundquist number, $h^{-1}$ is the inverse of the grid size, $\delta v$ is the amplitude of the random noise (RN) perturbation, $k$ is the wavenumber of the multi-mode (MM) perturbation, \giovani{and $t_{\rm max}$ is the time lasted by the simulation (in units of $t_A$)}. All models have magnetic Prandtl number $\text{Pr}_m \equiv \nu/\eta = 1$, and the initial thickness $\delta = S^{-1/2}$.}
    \label{tab:list_models}
\end{table}

\section{Measuring the reconnection rate}\label{sec:recrate}

We adapt the method described by \cite{kowal2009numerical} \citep[see also][]{Vicentin_2025} to calculate the reconnection rate from the unsigned magnetic flux. Specifically, we integrate $|B_x|$ over $x=0$, perpendicular to the current sheet. Since the signed flux through this plane is zero, dividing the unsigned integral by two gives the flux contribution from each polarity. As time evolves, the two initial flux ropes merge, and $|B_x|$ decreases due to the reconnection process. Then, we have the reconnection rate given by the time derivative of the unsigned magnetic flux:

\begin{equation}
    V_{\text{rec}} = - \frac{1}{2 |B_{x,0}|} \frac{\partial \Phi_B}{\partial t}= - \frac{1}{2 |B_{x,0}|} \frac{\partial}{\partial t}\int_{-0.5}^{0.5} |B_x| \, dy, \label{eq:vrec_kowal09_thiswork}
\end{equation}

\noindent where $|B_{x,0}|$ represents the initial amplitude of the non-reconnecting field, that in this configuration is given by $|B_{x,0}| = \max (|B_x|_{x=0}) \sim 1$, and $\Phi_B$ is the unsigned magnetic flux integrated across the center of the box ($x = 0$).

By adopting this method, we avoid problems caused by the accumulation of magnetic flux along the $x-$boundaries, since we are integrating the flux across the center of the box at $x=0$.\footnote{This choice differs from the method adopted by \cite{bhattacharjee2009fast}, where the reconnection rate is inferred from the maximum flux function in the reconnection layer. Here, instead, we compute the rate from the time derivative of the unsigned magnetic flux evaluated along the symmetry line perpendicular to the current sheet.}

\section{Results} \label{sec:results}

We discuss in this Section the results of 2D MHD simulations of current sheets for different values of Lundquist number, numerical grid resolution, and the two drivers of the initial perturbation. We also show a new method to compute the numerical error from the simulations using the components of the magnetic energy density equation.

\subsection{Models with initial random noise}

In the absence of external forcing or instabilities, the reconnection rate measured within the current sheet should follow the Sweet--Parker dependence on the Lundquist number, $V_{\text{rec}} \sim S^{-1/2}$, at least until a critical Lundquist number $S_c$ where the plasmoid instability can occur, making the reconnection rate reach a constant value, as verified, e.g., in the MHD simulations of \cite{bhattacharjee2009fast, huang2010scaling, Loureiro2012}. As emphasized previously, in these works, the authors found $S_c \sim 10^{4}$, and a ``universal'' rate of $V_\text{rec}/V_A \sim S_c^{-1/2} \sim 0.01$.

\begin{figure}[ht]
    \centering
    \includegraphics[width=0.99\linewidth]{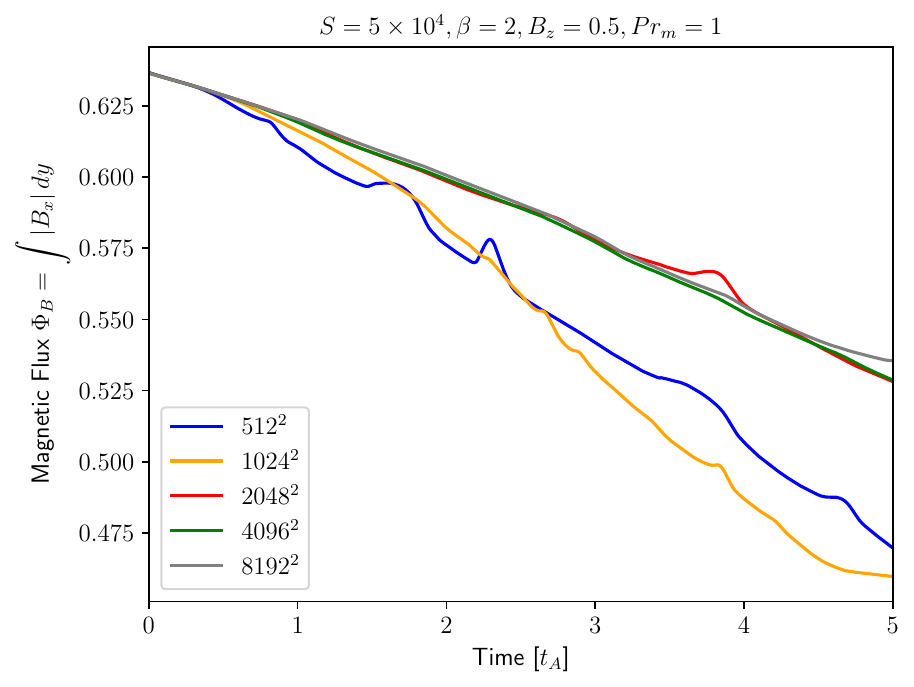}
    \caption{Time evolution of the magnetic flux $\Phi_B$ for 2D MHD simulations with initial random noise perturbation, $S=5\times 10^4$, $\delta v = 10^{-2} \, V_A$ and different resolutions. }
    \label{fig:Bflux_S5e4_resolutions}
\end{figure}

\begin{figure*}[ht!]
    \centering
    \includegraphics[width = 0.9 \textwidth]{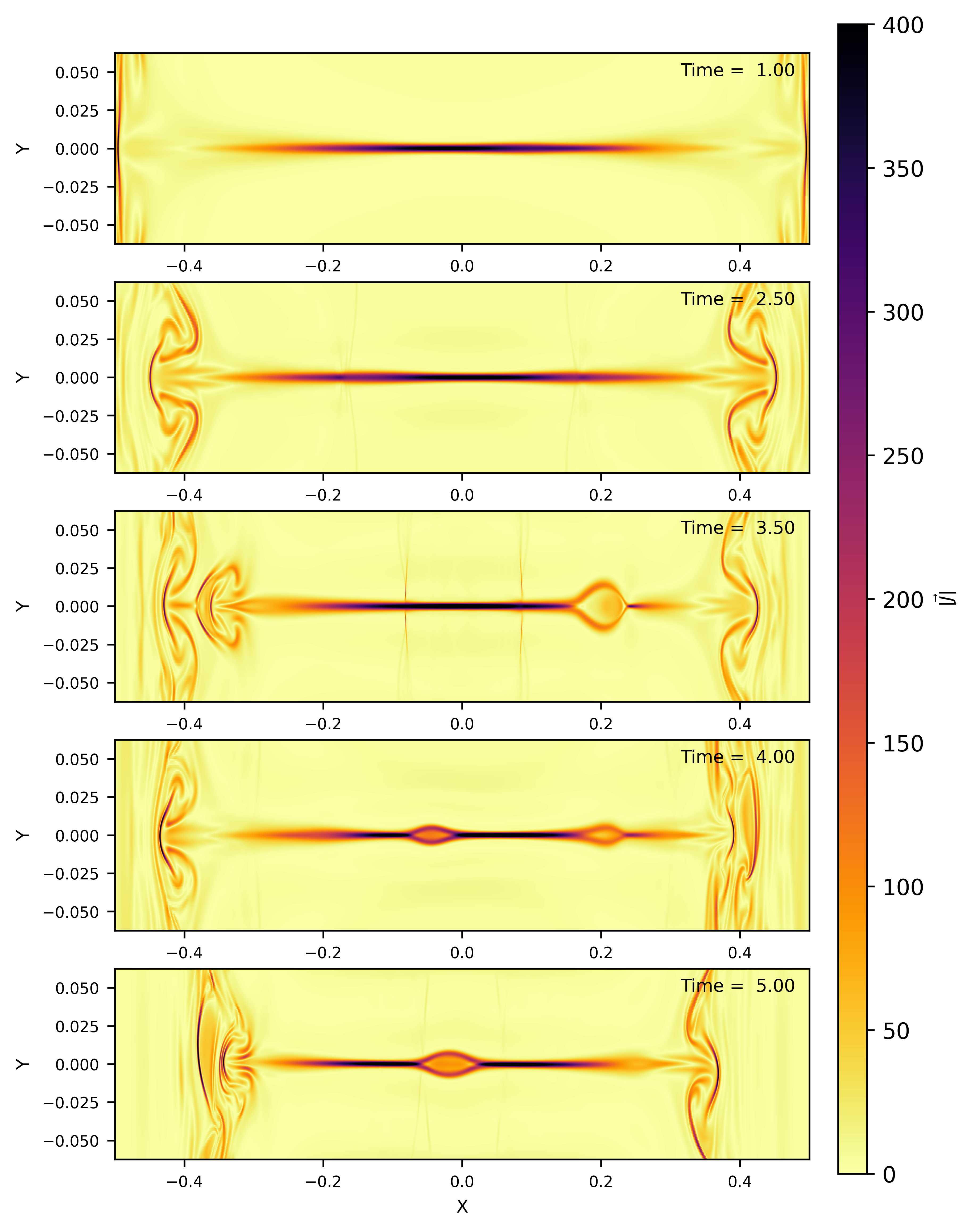}
    \caption{Colormaps of the current density magnitude ($|\mathbf{J}| = |\nabla \times \mathbf{B}|$) at different time-steps for the simulation with $S=5 \times 10^4$ and $h^{-1}=8192$. }
    \label{fig:cmaps_plasmoids_S5e4_h8k}
\end{figure*}

\begin{figure*}[ht!]
    \centering
    \includegraphics[width = 0.9 \textwidth]{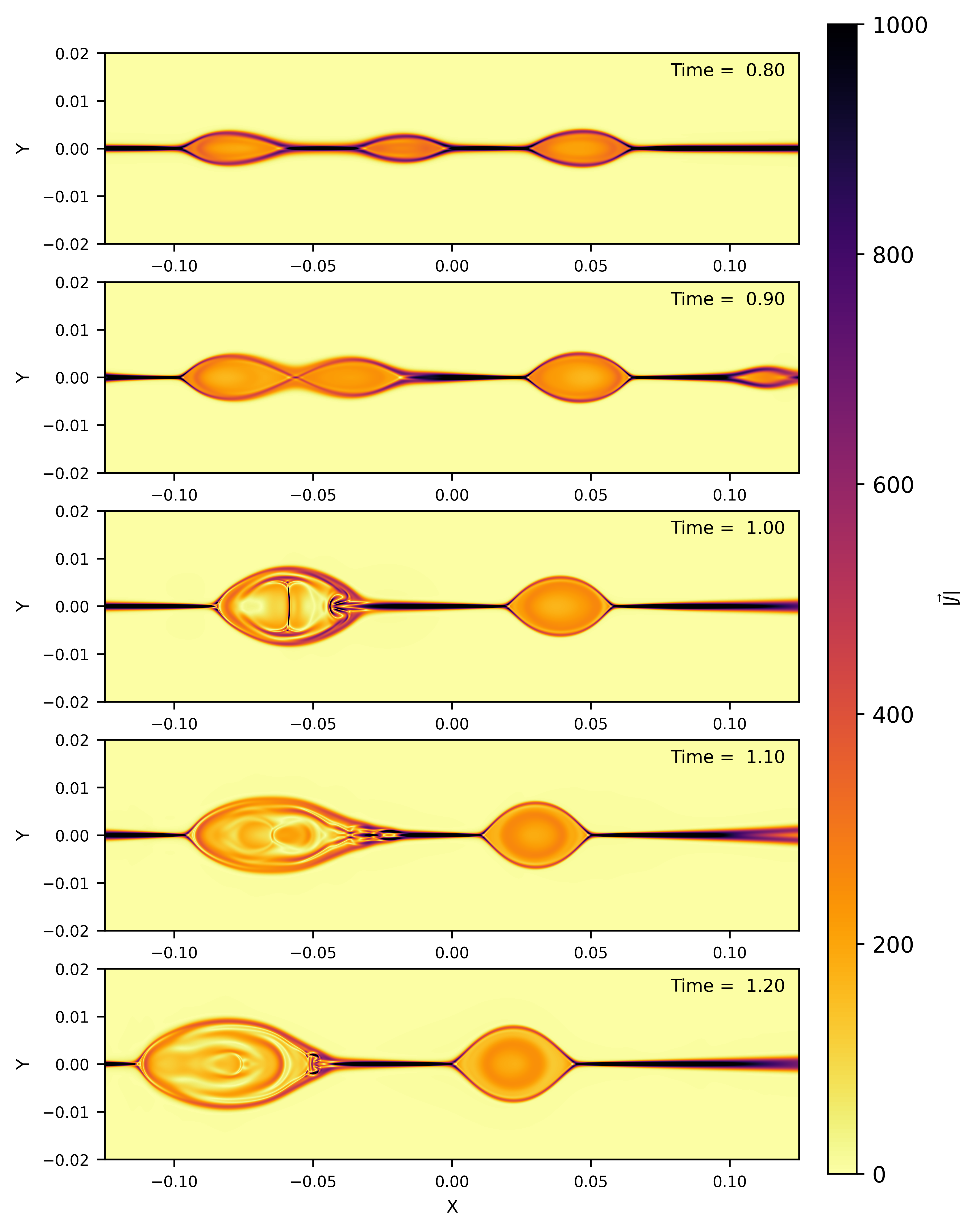}
    \caption{\giovani{Colormaps of the current density magnitude ($|\mathbf{J}| = |\nabla \times \mathbf{B}|$) at different time-steps for the simulation with $S=3.3 \times 10^5$ and $h^{-1}=65536$. In this case there is a   growing and merging of plasmoids, indicating a transition to the nonlinear regime of the tearing-mode instability.}}
    \label{fig:cmaps_plasmoids_S33e5_h65k}
\end{figure*}

In this work, for each Lundquist number tested in our simulations, we analyzed the convergence of magnetic flux over time for different grid resolutions. In Fig. \ref{fig:Bflux_S5e4_resolutions} we show the time evolution of the magnetic flux for  simulations with $S=5 \times 10^4$, using grid resolutions from $512^2$ to $8192^2$  cells.

From Fig. \ref{fig:Bflux_S5e4_resolutions}, we observe that convergence is achieved at a resolution $2048^2$. In this case, the tearing-mode instability still develops, and the bump seen at time $t \lesssim 4.0 \, t_A$ is attributed to the passage of a plasmoid through the center of the box ($x=0$). For higher resolutions, of $4096^2$ and $8192^2$, 
%even noticing the development of
although plasmoids still form, they are rapidly advected out of the domain (see Fig. \ref{fig:cmaps_plasmoids_S5e4_h8k}  for the resolution $h^{-1} = 8192$) and therefore do not affect the magnetic flux.

We note that, for the simulations shown in Fig. \ref{fig:Bflux_S5e4_resolutions},
the Lundquist number is $S = 5 \times 10^4$, which yields a current layer thickness of $\delta \sim S^{-1/2} \approx 4.47 \times 10^{-3}$. \giovani{With a grid resolution of $h^{-1} = 2048$ in
%to \giovani{resolve} 
the reconnection region, we obtain a ratio of $\delta / h \approx 9.15$. 
The convergence criterion proposed by \cite{Morillo2025}, $\delta/h \geq 10$, therefore provides a useful guideline for evaluating numerical resolution in our run with $2048^2$ and higher grids. However, in contrast to their results, we still observe plasmoid formation even in cases with $\delta/h > 10$.}

\giovani{In Appendix \ref{Appendix:Minimum_Resolution}, we estimate the minimum resolution needed to resolve the current sheet in the tearing-mode 
%plasmoid-mediated 
reconnection regime. For the range of $S$ explored here, our analysis supports $\delta/h \gtrsim 10$ as a reasonable choice. We note, however, that the minimum required number of grid cells also depends on the Lundquist number, scaling as $n \propto S^{1/8}$.}

For example,  Figure \ref{fig:cmaps_plasmoids_S5e4_h8k} shows 2D colormaps of the current density magnitude ($|\mathbf{J}| = |\nabla \times \mathbf{B}|$) at different snapshots of the simulation with $S = 5 \times 10^4$ and $h^{-1} = 8192$. In this converged regime, plasmoids are generated but are rapidly advected out of the 2D domain without merging or evolving into ``monster'' plasmoids.\footnote{It is worth noting that, \giovani{in all of our simulations, we observe waves propagating from the domain boundaries toward the center (see, for example, the snapshot at $t = 3.5\, t_A$ in Fig.~\ref{fig:cmaps_plasmoids_S5e4_h8k}). This behavior arises from the imposed closed boundary conditions and from compressibility effects, since we adopt a moderate plasma beta, $\beta = 2$, which differs from previous studies \citep[e.g.,][where $\beta \ge 6$]{bhattacharjee2009fast, huang2010scaling, Huang_2017}}. These waves can influence the system by introducing additional perturbations, which in turn may contribute to plasmoid formation.}
Even when the simulation is extended to much longer timescales, the absence of growing plasmoids remains, as demonstrated in Figure \ref{fig:curdens_S1e5_upt10} in the Appendix \ref{sec:appendix_sim} for a run lasting up to $t_{\rm max} =10 \, t_A$ with $S = 10^5$.

\giovani{
The behavior described above persists up to $S = 2 \times 10^5$. Figure~\ref{fig:cmaps_plasmoids_S33e5_h65k} shows the time evolution of the 2D colormap of the current-density magnitude for a simulation with $S = 3.3 \times 10^5$ on a uniform grid with $h^{-1} = 65536$, corresponding to $\delta/h \sim 113$. In this case, we clearly observe plasmoids growing and subsequently merging, indicating a transition to the nonlinear development of the tearing-mode instability. In the following diagrams, we quantify these different regimes more systematically.}

\begin{figure}[ht]
    \centering
    \includegraphics[width=0.99\linewidth]{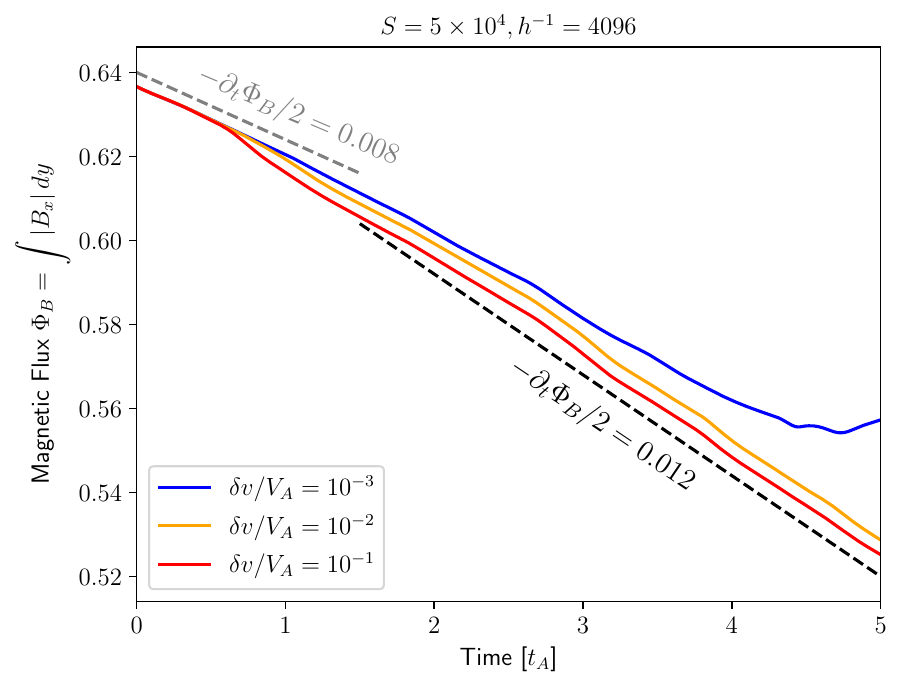}
    \caption{Time evolution of the magnetic flux $\Phi_B$ for simulations with initial random noise perturbation, $S=5\times 10^4$, $h^{-1} = 4096$, and different amplitudes of noise $\delta v$. 
     }
    \label{fig:deltaV_S5e4}
\end{figure}

Figure \ref{fig:deltaV_S5e4} shows the time evolution of the magnetic flux for simulations with $S = 5 \times 10^4$, resolution $h^{-1} = 4096$, and different amplitudes of the initial random velocity perturbation, $\delta v = \{10^{-3}, 10^{-2}, 10^{-1}\} \, V_A$. The initial reconnection rate is $V_{\text{rec}}/V_A = 0.008$ in all cases. The differences arise from the amplitude of the perturbation: for $\delta v/V_A = 10^{-1}$ (red curve), the first plasmoids appear at $t \sim 0.6 \, t_A$, after which the reconnection rate increases to $0.012$. For $\delta v/V_A = 10^{-2}$ (orange), this transition occurs at $t \sim 1.0 \, t_A$, and for $\delta v/V_A = 10^{-3}$ (blue) it is delayed until $t \sim 2.6 \, t_A$. Unless otherwise specified, we adopt $\delta v = 10^{-2} \, V_A$ as the standard amplitude of the initial random perturbation.

\begin{figure}[ht]
    \centering
    \includegraphics[width=0.99\linewidth]{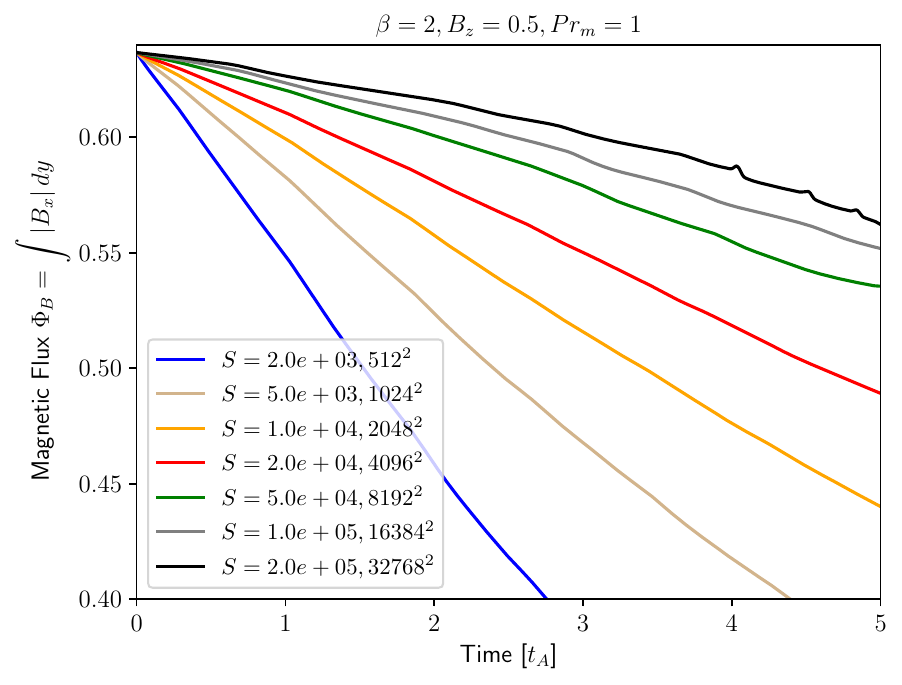}
    \caption{Time evolution of the magnetic flux $\Phi_B$ for simulations with initial random noise perturbation and different Lundquist numbers \giovani{between $S =2 \times 10^3$ and $S=2\times 10^5$}, at their respective best resolutions. \giovani{For this range of $S$ values, the nonlinear regime of the tearing-mode instability is not achieved.} 
    }
    \label{fig:depS_well_resolved}
\end{figure}

By plotting the time evolution of the magnetic flux for simulations with different Lundquist numbers in the range $S = 2 \times 10^3$ to $2 \times 10^5$, for which convergence is achieved (i.e., $\delta/h > 10$) \giovani{and no plasmoid merging occurs,} we obtain Figure ~\ref{fig:depS_well_resolved}.
It should be noted that, for $S = 2 \times 10^5$ (black curve), the resolution of $h^{-1} = 32768$ \giovani{also} resolves the current sheet, \giovani{ since $\delta \approx 2.24 \times 10^{-3}$ satisfies the threshold condition, where $\delta / h  \approx 72$}. \giovani{However, in} this case, plasmoid passages interfere with the computed magnetic flux -- evidenced by the \giovani{very} small bumps in the black curve -- and consequently \giovani{minimally} affect the inferred reconnection rate. Notably, as the Lundquist number increases and plasmoid formation becomes more frequent, plasmoids crossing the line $x = 0$ can produce more frequent negative instantaneous derivatives in the measured reconnection rate.

\begin{figure}[h!]
    \centering
    \includegraphics[width=0.95 \linewidth]{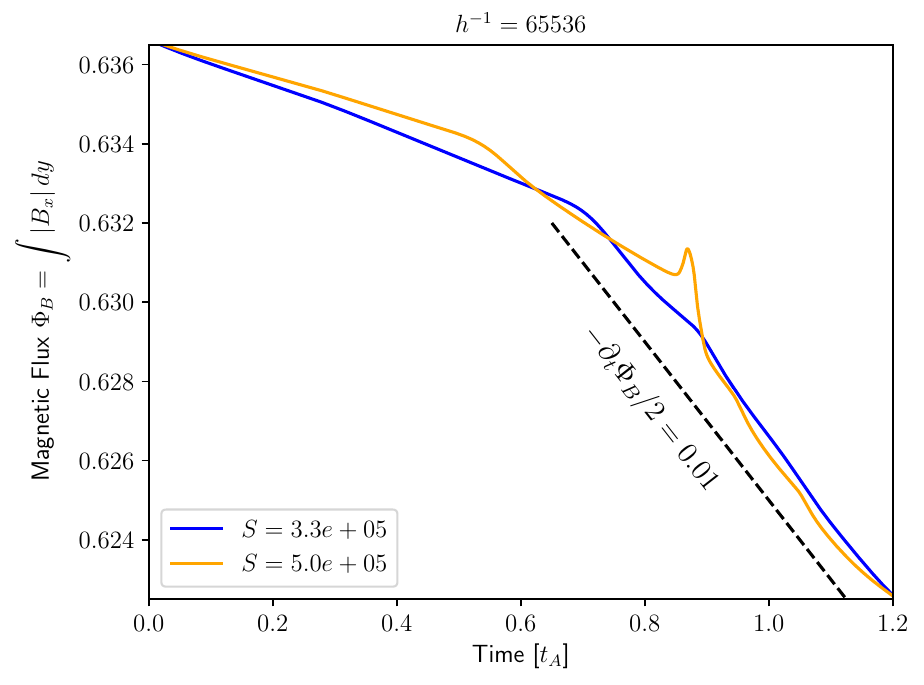}
    \caption{\giovani{Time evolution of the magnetic flux for the simulations with $S=3.3 \times 10^5$ (blue) and $5\times 10^5$ (yellow), both with $h^{-1} = 65536$, for which the nonlinear regime of tearing instability is achieved and the reconnection rate saturates at $V_{\rm rec} \sim 0.01 \, V_A$ (black dashed line).}}
    \label{fig:bflux_tearing_S3e5_5e5}
\end{figure}

\begin{figure}[h!]
    \centering
    \includegraphics[width=0.95 \linewidth]{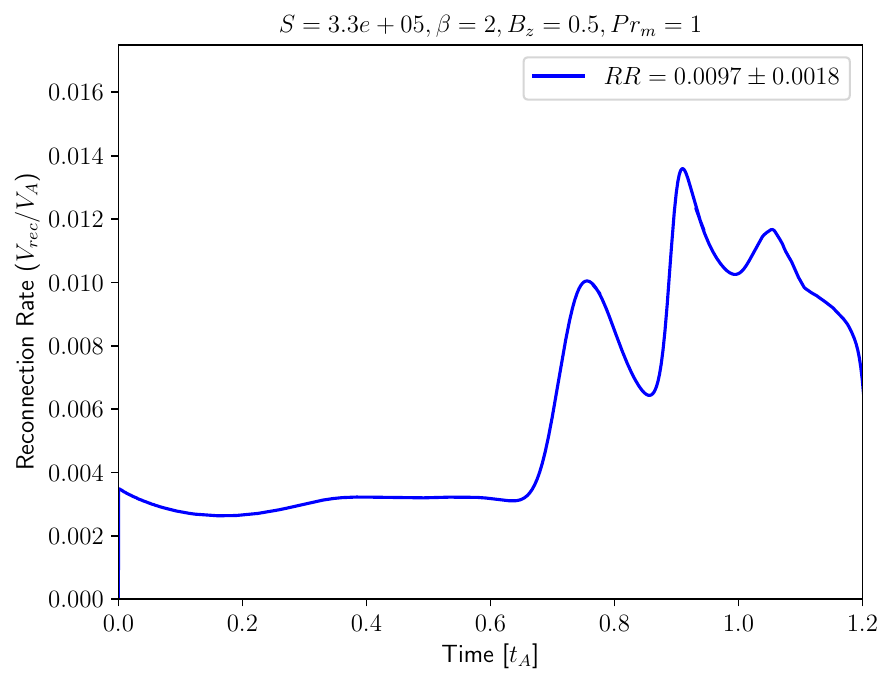}
    \caption{\giovani{Time evolution of the reconnection rate for the simulation with $S=3.3 \times 10^5$ and $h^{-1} = 65536$, where the nonlinear regime of the  tearing instability is achieved and the reconnection rate saturates at $V_{\rm rec} \sim 0.01 \, V_A$. The average was taken between $0.7 \le t \le 1.2 \, t_A$.}}
    \label{fig:vrec_error_S3e5}
\end{figure}

\giovani{Complementing Figure~\ref{fig:depS_well_resolved}, Figure~\ref{fig:bflux_tearing_S3e5_5e5} shows the time evolution of the magnetic flux for two other simulations that reach the nonlinear tearing-mode regime, with $S=3.3\times10^5$ and $5\times10^5$. In the latter case, the uniform grid also has $h^{-1}=65536$, corresponding to $\delta/h\sim92$. For these high-$S$ runs, we observe the onset of nonlinear tearing at $t\sim0.6,t_A$ and a rapid saturation of the reconnection rate at $V_\text{rec}/V_A\sim0.01$ (see also Figure~\ref{fig:vrec_error_S3e5}).}

\begin{figure*}[ht]
    \centering
    \includegraphics[width = 0.75 \textwidth]{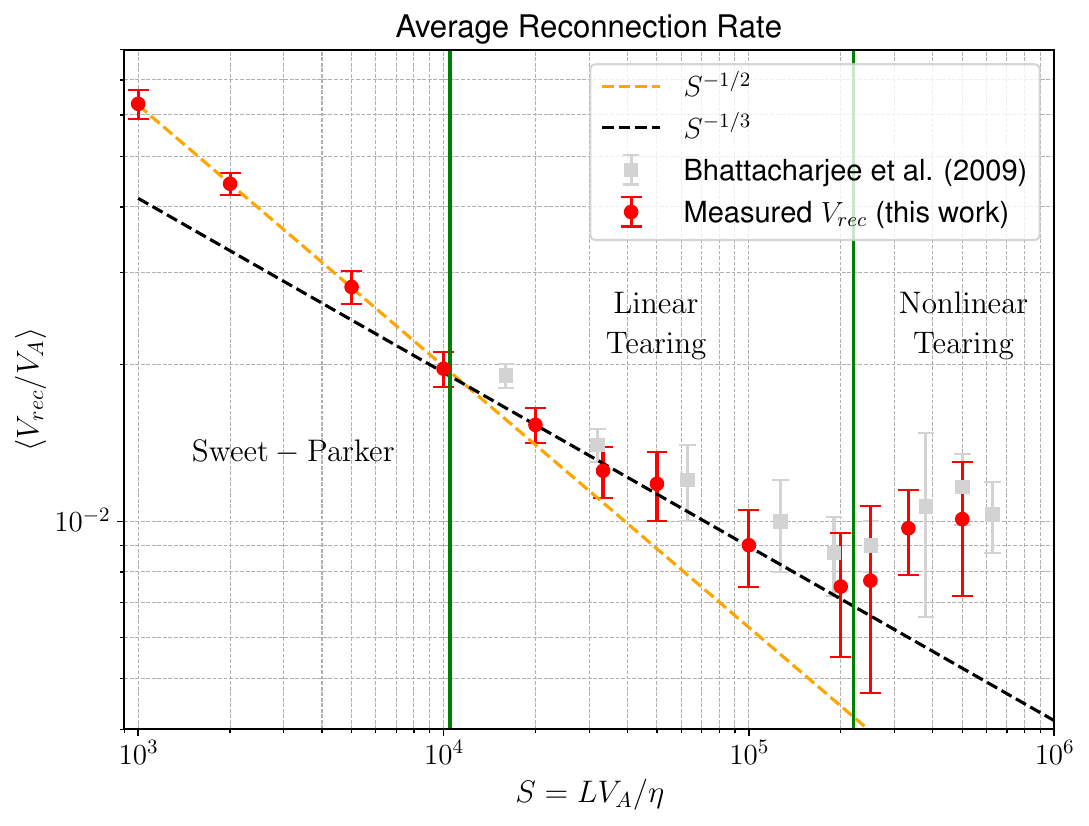}
    \caption{Dependence of the average reconnection rate $V_{\text{rec}}$ on the Lundquist Number $S$ for models with initial random noise perturbation. The dashed orange line represents the Sweet--Parker scaling of $V_{\text{rec}} \propto S^{-1/2}$, and the black dashed line represents the scaling for the linear tearing mode-mediated reconnection $V_{\text{rec}} \propto S^{-1/3}$ -- as in Eq. (\ref{eq:Vrec_max}). \giovani{The vertical lines separate the three reconnection regimes, namely, the Sweet--Parker regime which persists up to  $S \sim 10^4$; the linear tearing-mode regime, occurring for   $2\times 10^4\lesssim S \lesssim 2 \times 10^5$, where the reconnection rate exhibits a modest increase, $V_{\text{rec}} \sim S^{-1/3}$, yet remains resistivity-dependent; and the nonlinear tearing-mode regime for 
    $S > 2 \times 10^5$, where the reconnection rate saturates at $V_\text{rec} / V_A \sim 0.01$. These results contradict earlier claims that tearing-mode reconnection becomes fast and resistivity-independent for $S \gtrsim 10^4$. For comparison, we overlay the numerical results from the simulations of \citet{bhattacharjee2009fast} (light-gray symbols). Their values in the range $10^4 \lesssim S \lesssim 2 \times 10^5$ align more closely with a slow, linear tearing-mode regime—consistent with our findings—rather than with a fast reconnection \giovani{$S-$independent} regime in this range of Lundquist numbers.}}
    \label{fig:Average_Vrec_depS-13}
\end{figure*}

Subsequently, by taking the time derivative of the curves in Figures \ref{fig:depS_well_resolved} and \ref{fig:bflux_tearing_S3e5_5e5}, and normalizing according to Eq. (\ref{eq:vrec_kowal09_thiswork}), we obtain the average reconnection rates as a function of the Lundquist number, shown in Figure \ref{fig:Average_Vrec_depS-13}.
%Fig.~\ref{fig:Average_Vrec_depS}.

In Fig. \ref{fig:Average_Vrec_depS-13},
%Fig.~\ref{fig:Average_Vrec_depS}, 
the simulations with $S = 10^3$ and $S = 2 \times 10^3$ were evolved only up to $t_\text{max} = 3.0 \, t_A$, as the magnetic flux decays rapidly and the two flux tubes quickly coalesce into a single structure. This coalescence reduces the reconnection rate at later times. For these cases, we averaged the reconnection rate over the interval $0 \leq t \leq 2.0 \, t_A$. For $S = 5 \times 10^3$, we performed the averaging 
%was performed 
over $1.0 \, t_A \leq t \leq 3.0 \, t_A$, while for all higher-$S$ simulations \giovani{in which the tearing instability develops, the averaging window corresponds to the stage where the plasmoid activity produces a statistically steady reconnection rate, which corresponds to times from the onset of the plasmoid formation up to $t_{\rm max}$ (see Table \ref{tab:list_models}), except for the case of $S = 2\times 10^5$, since the interval $4 - 5 \, t_A$ is dominated by large, abrupt oscillations caused by the advection of plasmoids across $x = 0$, which drives $\Phi_B(t)$ through local extrema and can yield spurious negative instantaneous derivatives. For this run, a more stable linear region is observed between $t = 2.0 \, t_A$ and $t = 4.0 \, t_A$.
}

From Fig. \ref{fig:Average_Vrec_depS-13},
%\ref{fig:Average_Vrec_depS}, 
we observe a good agreement between the measured reconnection rates and the Sweet--Parker scaling, $V_\text{rec}/V_A \sim S^{-1/2}$ (orange dashed line) for the Lundquist numbers up to $S \sim 10^4$. For higher values, \giovani{$2\times 10^4\lesssim S \lesssim 2 \times 10^5$,} the reconnection rates deviate from the slow Sweet--Parker regime due to the development of \giovani{the tearing-mode instability}. In this regime, the rates are slightly enhanced but still depend on the Ohmic resistivity and Lundquist number, following $V_\text{rec}/V_A \sim S^{-1/3}$ (black dashed line), and thus remain slow. Remarkably, this 
$S^{-1/3}$ scaling is the same as the one obtained in Section \ref{Sec:Linear_inst_TM} (Eq. \ref{eq:Vrec_max}) from linear theory of reconnection driven by tearing-mode.\footnote{We also note that the difference between the $S^{-1/3}$ scaling  and the one obtained in  \citealt{lazarian1999reconnection}, $S^{-3/10}$  (see eq. \ref{eq: vrec_LV}), is insignificant and well within the numerical errors.}

\giovani{In practice, however, we only observe clear plasmoid formation in runs with $S \ge 3.3 \times 10^4$. For $S=2\times 10^4$, the reconnection rate increases only modestly, from the initial Sweet--Parker value $V_{\rm rec}/V_A \sim 0.0140$ to an average $V_{\rm rec}/V_A = 0.0153(12)$. This mild enhancement may indicate the development of a  marginally unstable tearing regime. In this case, tearing modes grow but have not yet produced long-lived, macroscopic plasmoids.}

\giovani{In Fig. \ref{fig:Average_Vrec_depS-13}, the orange dashed line shows the best-fit scaling for the Sweet--Parker regime, $V_{\rm rec,SP}=\alpha_{SP}S^{-1/2}$, while the black dashed line shows the best-fit scaling for the linear tearing-mode regime, $V_{\rm rec,TM}=\alpha_{TM}S^{-1/3}$. The fitted prefactors are $\alpha_{SP}=1.9813$ and $\alpha_{TM}=0.4153$. We can then estimate the critical Lundquist number for the transition between these regimes by equating the two fits, $V_{\rm rec,SP}(S_c)=V_{\rm rec,TM}(S_c)$, which yields}

\begin{equation}
    \giovani{S_c = \left( \dfrac{\alpha_{SP}}{\alpha_{TM}} \right)^6 \approx 1.2 \times 10^4,}
\end{equation}

\noindent
\giovani{corresponding to the leftmost vertical green line in Fig. \ref{fig:Average_Vrec_depS-13}.}

\giovani{Only for $S >2\times 10^5$ does the system reach the $nonlinear$ stage of the tearing-mode instability, where the reconnection rate saturates at $V_\text{rec}/V_A \sim 0.01$. This result contradicts earlier studies \citep[e.g.,][]{bhattacharjee2009fast, Loureiro2012} that predicted such saturation at values of $S$ roughly an order of magnitude lower ($S \sim 10^4$). 
\giovani{Clearly, for $10^4 \lesssim S \lesssim 2\times 10^5$, the system lies in the linear tearing-mode regime, where reconnection  is slow and still resistive-dependent, with $V_{\text{rec}} \sim S^{-1/3}$.}
For comparison, Fig.~\ref{fig:Average_Vrec_depS-13} also shows the simulation results of \citet{bhattacharjee2009fast}, which further highlight this contradiction -- notice that their data points (light-gray squares) for $10^4 < S  \lesssim 2 \times 10^5$ also fit the relation $V_{\rm rec} \propto S^{-1/3}$ in the ``Linear Tearing'' range.}

We emphasize that, in this work, ``linear" and ``nonlinear" designate operational regimes of the global evolution. The former corresponds to cases where the reconnection rate follows the scaling predicted by linear tearing theory and in which plasmoids are advected away before strong interaction occurs. The latter corresponds to cases in which plasmoids are generated faster than they are expelled, allowing interaction and coalescence and leading to nonlinear plasmoid dynamics.

\subsection{Models with multi-mode small-scale initial perturbation}

As pointed out earlier, we adopted two different methods to drive the initial perturbation into the 2D domain. In this section, we discuss the results of the simulations with multi-mode, small-scale ($k \gg 1$) initial perturbation \citep[see, e.g.,][]{alvelius1999random, kowal2009numerical, Kulpa-Dybel-2010}.

\begin{figure}[ht]
    \centering
    \includegraphics[width=0.99\linewidth]{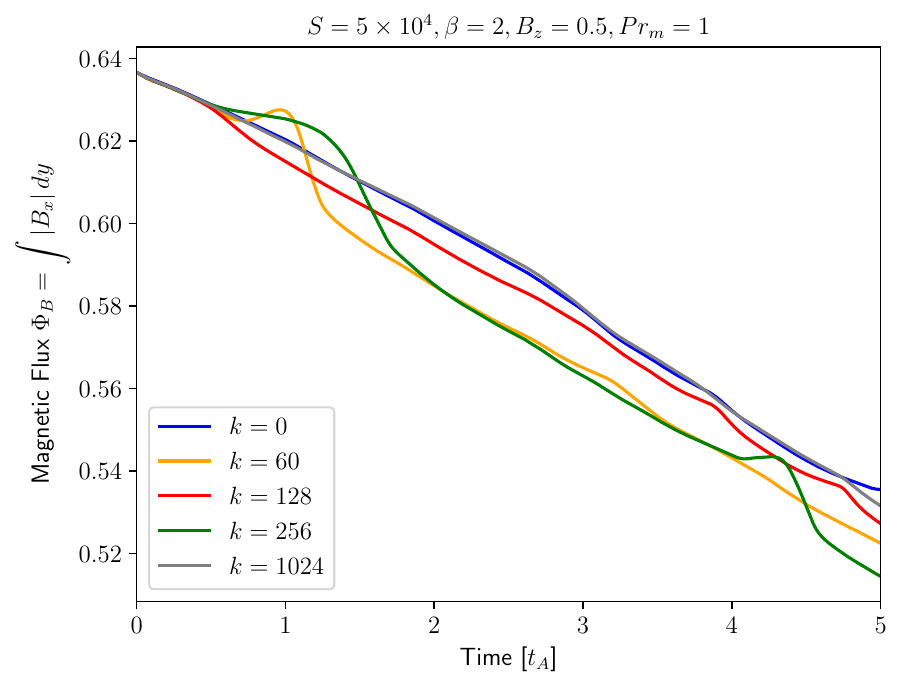}
    \caption{Magnetic flux evolution for the simulations 
    with initial small-scale perturbation, $S = 5 \times 10^4$ and resolution $h=1/8192$. The colors represent different wavenumbers of the initial perturbation, injected from $t=0$ up to $0.1 \, t_A$. The only exception is the  blue curve, which  was obtained using initial random noise as in Figures \ref{fig:Bflux_S5e4_resolutions} to \ref{fig:Average_Vrec_depS-13}, and is shown here for comparison.}
    \label{fig:S5e4_Diff_k_injs}
\end{figure}

In Fig.~\ref{fig:S5e4_Diff_k_injs}, we present the time evolution of the magnetic flux for the simulations with $S = 5 \times 10^4$ and $h^{-1}=8192$. 
Except for the blue curve, which corresponds to a model initialized with random noise (as in
Fig. \ref{fig:depS_well_resolved}), all other simulations in Fig. \ref{fig:S5e4_Diff_k_injs} were perturbed using small-scale, multi-mode fluctuations peaked around a wavenumber $k$ and injected over a short interval (from $t=0$ up to $0.1 \, t_A$). We find that perturbations at scales of $\ell \sim k^{-1}$ are close to the resistive scale of the simulation, making them effective at triggering small-scale instabilities such as the tearing mode. The modes satisfy $ |\mathbf{k}| = k$, but with the constraint $k_x < (2 \pi \delta)^{-1}$, where $\delta$ is the initial current sheet thickness.

Despite small variations in the curves, primarily caused by the passage of plasmoids through the center of the domain, all runs are consistent with a nearly constant reconnection rate of $V_\text{rec}/V_A \sim 0.012$, comparable to the value obtained for the model initialized with random noise (blue curve, see also Figure 
 \ref{fig:Average_Vrec_depS-13}).
%\ref{fig:Average_Vrec_depS}). 
This result is also in agreement with the scaling $V_\text{rec} \propto S^{-1/3}$ predicted for tearing mode-driven reconnection.

\begin{figure}[ht]
    \centering
    \includegraphics[width=0.99\linewidth]{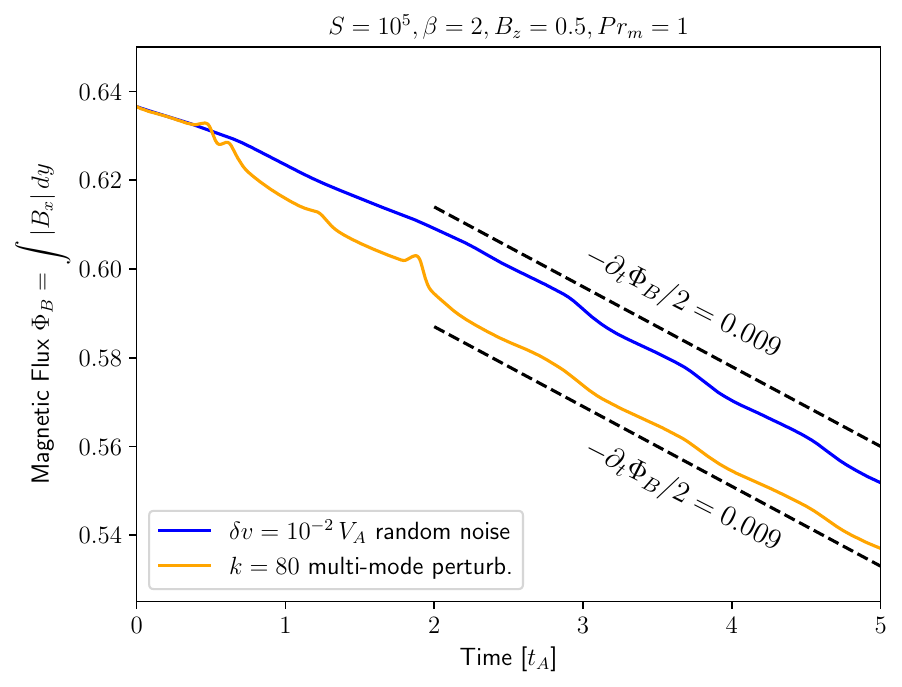}
    \caption{Magnetic flux evolution for the simulations with initial random noise ($\delta v = 10^{-2} \, V_A$, blue) and small-scale ($k = 80$, orange) perturbation, 
    $S = 10^5$ and resolution $h=1/16384$. The dashed black line corresponds to $V_\text{rec} = 0.009 \, V_A$, the average reconnection rate measured in $2.0 \le t \le 5.0 \, t_A$. 
    }
    \label{fig:S1e5_Diff_k_injs}
\end{figure}

In Fig. \ref{fig:S1e5_Diff_k_injs}, we present the same analysis for simulations with $S=10^5$ and two initial perturbations: random noise with $\delta v = 10^{-2} \, V_A$  (blue), and a multi-mode perturbation with $k=80$ (orange). In both cases the system reaches the same average reconnection rate  $\langle V_\text{rec}\rangle = 0.009 \, V_A$ which is also consistent with the \giovani{linear regime of}  tearing-mode-driven reconnection scale $V_\text{rec} \propto S^{-1/3}$.

\subsection{Error estimation}
\label{ssec:Error_Estimation}

Although significant efforts have explored magnetic reconnection in the high–Lundquist-number regime, systematic convergence studies and quantitative evaluations of numerical error remain scarce. If the increase of the Lundquist number $S$ is achieved by decreasing the resistivity $\eta$, the current sheet becomes extremely thin, with its initial laminar thickness scaling as $\delta \sim \eta^{1/2}\sim S^{-1/2}$,
under the Sweet--Parker model, thereby requiring very high spatial resolution for accurate representation. Without rigorous error quantification, it is challenging to determine whether reported results reflect genuine physical behavior or are affected by artifacts introduced by insufficient numerical resolution. In this section we present a  method to estimate the errors of our analysis.

\begin{figure*}[ht]
    \centering \includegraphics[width=0.487\linewidth]{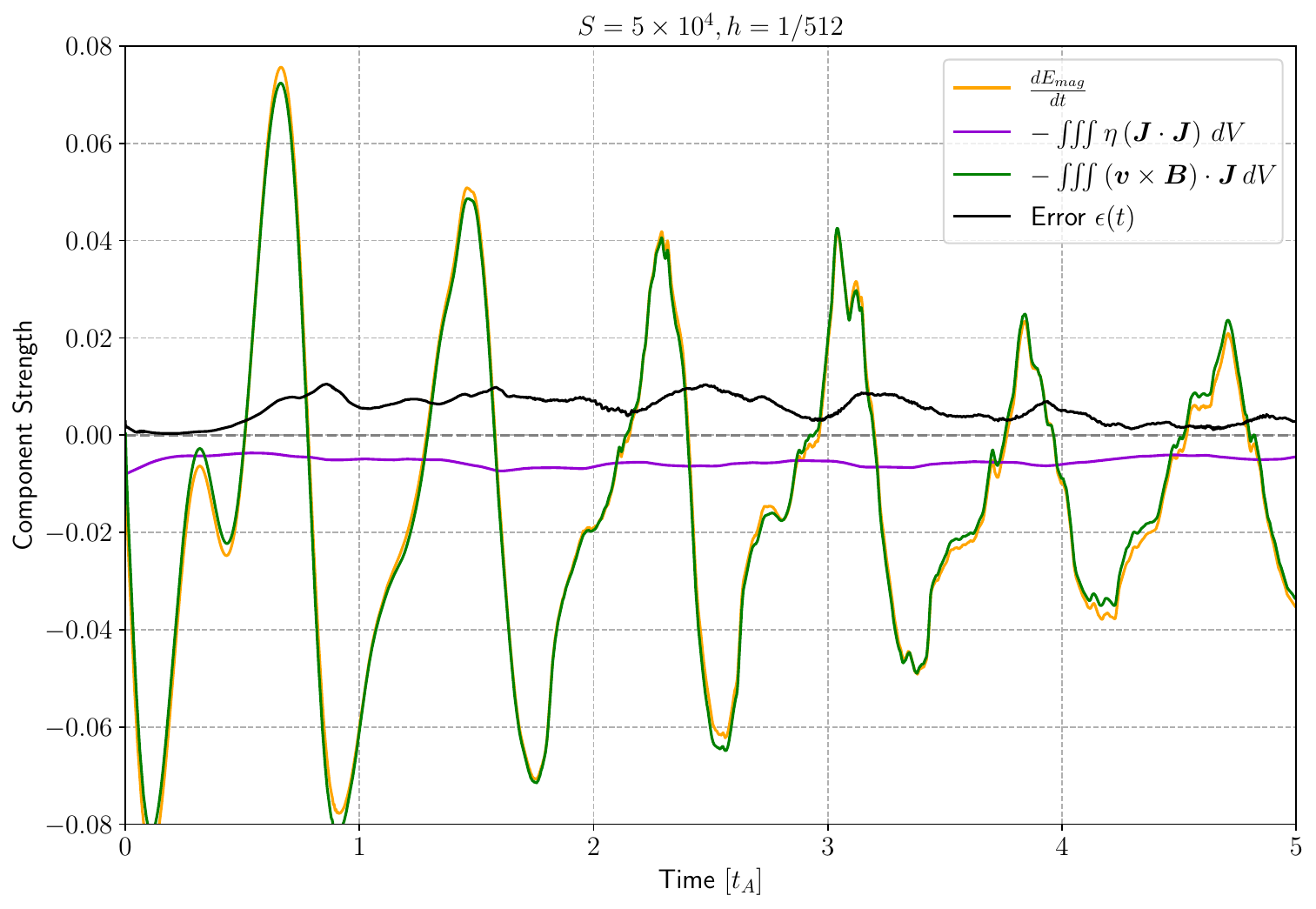}
    \includegraphics[width=0.49\linewidth]{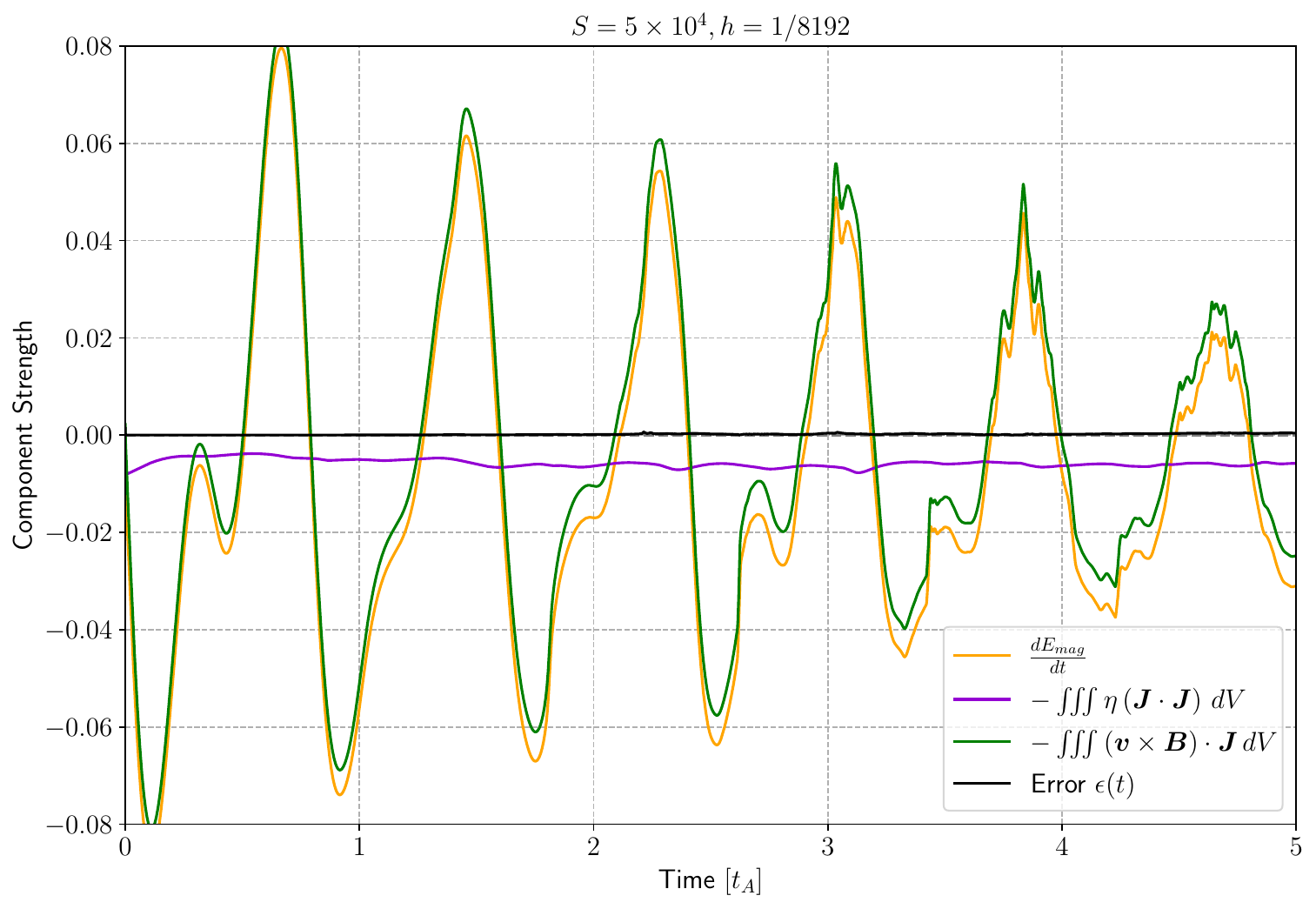}
    \caption{Components of the Eq. (\ref{eq:mag_energy_evol}) and numerical error estimation (black solid line) from Eq. (\ref{eq:error_mag_terms}) for the simulation with initial random noise, $S=5 \times 10^4$ and $h=1/512$ (left) and $h=1/8192$ (right).}
    \label{fig:error_S5e4_h512_h8k}
\end{figure*}

In resistive MHD, the time evolution of the magnetic energy density $E_B$ is governed by the volumetric integral equation:

\begin{align}
\frac{dE_B}{dt} &= - \iiint_V (\vec{v} \times \vec{B}) \cdot \vec{J} \, dV - \iiint_V \eta |\vec{J}|^2 \, dV \nonumber \\
&\quad + {\iint_{\partial V} \left[ \vec{E} \times \vec{B} \right] \cdot \hat{n} \, dS}, \label{eq:mag_energy_evol}
\end{align}

\noindent
where $ E_B \equiv \iiint_V \frac12 |\vec{B}|^2\, dV$ is the magnetic energy integrated across the entire simulated box, and the Poynting flux $\iint_{\partial V} \left[ \vec{E} \times \vec{B} \right] \cdot \hat{n} \, dS $ vanishes due to the boundary conditions imposed. Therefore, we can estimate the numerical error from the components of the Eq. (\ref{eq:mag_energy_evol}), by assuming

\begin{equation}
    \epsilon = \left| \frac{dE_B}{dt} + \iiint_V (\vec{v} \times \vec{B}) \cdot \vec{J} \, dV + \iiint_V \eta |\vec{J}|^2 \, dV  \right|. \label{eq:error_mag_terms}
\end{equation}

In Fig. \ref{fig:error_S5e4_h512_h8k} (left) we show the time evolution of all components of Eq. (\ref{eq:error_mag_terms}) along with  the estimated numerical error, $\epsilon$ (black solid line), for the simulation with $S = 5 \times 10^4$ and $h^{-1} = 512$. 
%We notice that
For this low resolution case, the numerical error is significant and sometimes even exceeds the amplitude of the heating term ($\eta \mathbf{J}^2$, purple line). This indicates that, at this resolution, the simulation is not well resolved, which is consistent with the convergence analysis presented in Fig. \ref{fig:Bflux_S5e4_resolutions}.

Repeating the same analysis for the higher resolution $h^{-1} = 8192$, we obtain Fig. \ref{fig:error_S5e4_h512_h8k} (right). In this case,  the numerical error $\epsilon$ is significantly reduced and remains smaller than each individual component of the magnetic terms (in magnitude), demonstrating that the simulation is well resolved at this resolution.

\begin{figure}[ht]
    \centering
\includegraphics[width=0.99\linewidth]{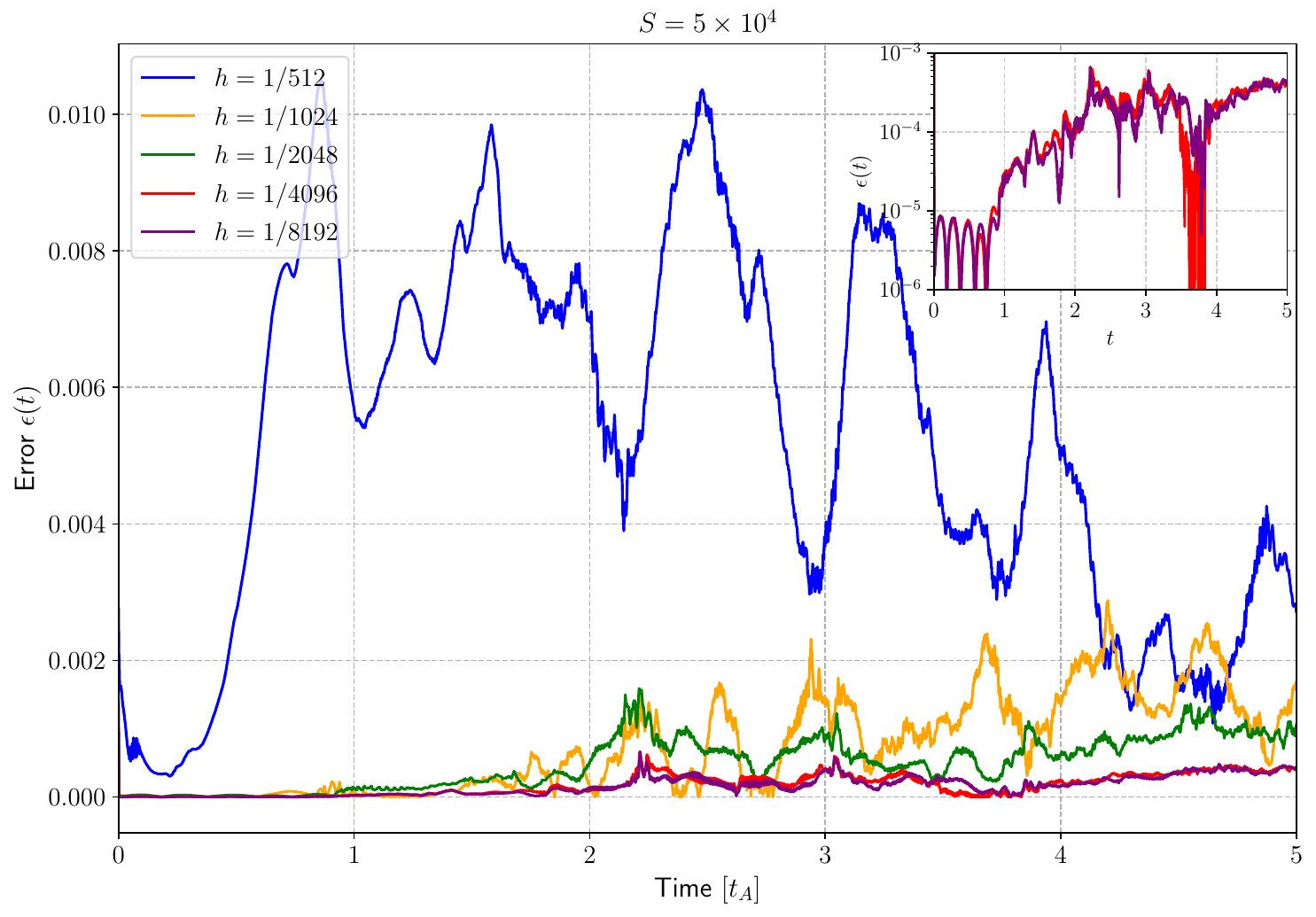}
    \caption{Numerical error estimate calculated from Eq. (\ref{eq:error_mag_terms}) for simulations with initial random noise, $S=5 \times 10^4$, and different grid resolutions. The inset plot shows the time evolution of the error $\epsilon (t)$ on a logarithmic scale for the two highest resolution simulations.}
    \label{fig:err0r_S5e4_dif_resol}
\end{figure}

 In Fig. \ref{fig:err0r_S5e4_dif_resol}, we present the numerical error estimate computed from Eq. (\ref{eq:error_mag_terms}) for simulations with $S=5 \times 10^4$ and different resolutions, ranging from  $h^{-1} = 512$ to $8192$. \giovani{In this case,  convergence is achieved for the highest resolutions of $h^{-1} = 4096$ and $8192$ (red and purple curves, respectively)}, where the condition $\delta / h > 18$ is satisfied, which is even more stringent than the threshold reported by \cite{Morillo2025}.

The quantity shown in Fig. \ref{fig:err0r_S5e4_dif_resol} is not a pure truncation-error measure of the MHD solver. Rather, it is the residual between the time derivative of the domain-integrated magnetic energy and the domain-integrated right-hand side of Eq. (\ref{eq:mag_energy_evol}), and therefore also reflects uncertainties introduced by the post-processing procedure used to evaluate this balance. In particular, the time derivative of the integrated magnetic energy is computed a posteriori with a lower-order operator than the one used in the simulation time integration. Consequently, once the spatial discretization error becomes sufficiently small, the residual is not expected to follow the formal convergence rate of the numerical scheme alone. In addition, the use of cell-averaged quantities in the evaluation of both the magnetic energy and the terms in Eq. (\ref{eq:mag_energy_evol}) introduces a small intrinsic mismatch in the discrete balance. At high resolution, the residual approaches a floor consistent with the $10^{-4}$ absolute and relative tolerances of the embedded time integrator, which explains the saturation of the two highest-resolution curves (see the inset plot in Fig. \ref{fig:err0r_S5e4_dif_resol}).

\begin{figure}[ht]
    \centering
\includegraphics[width=0.99\linewidth]{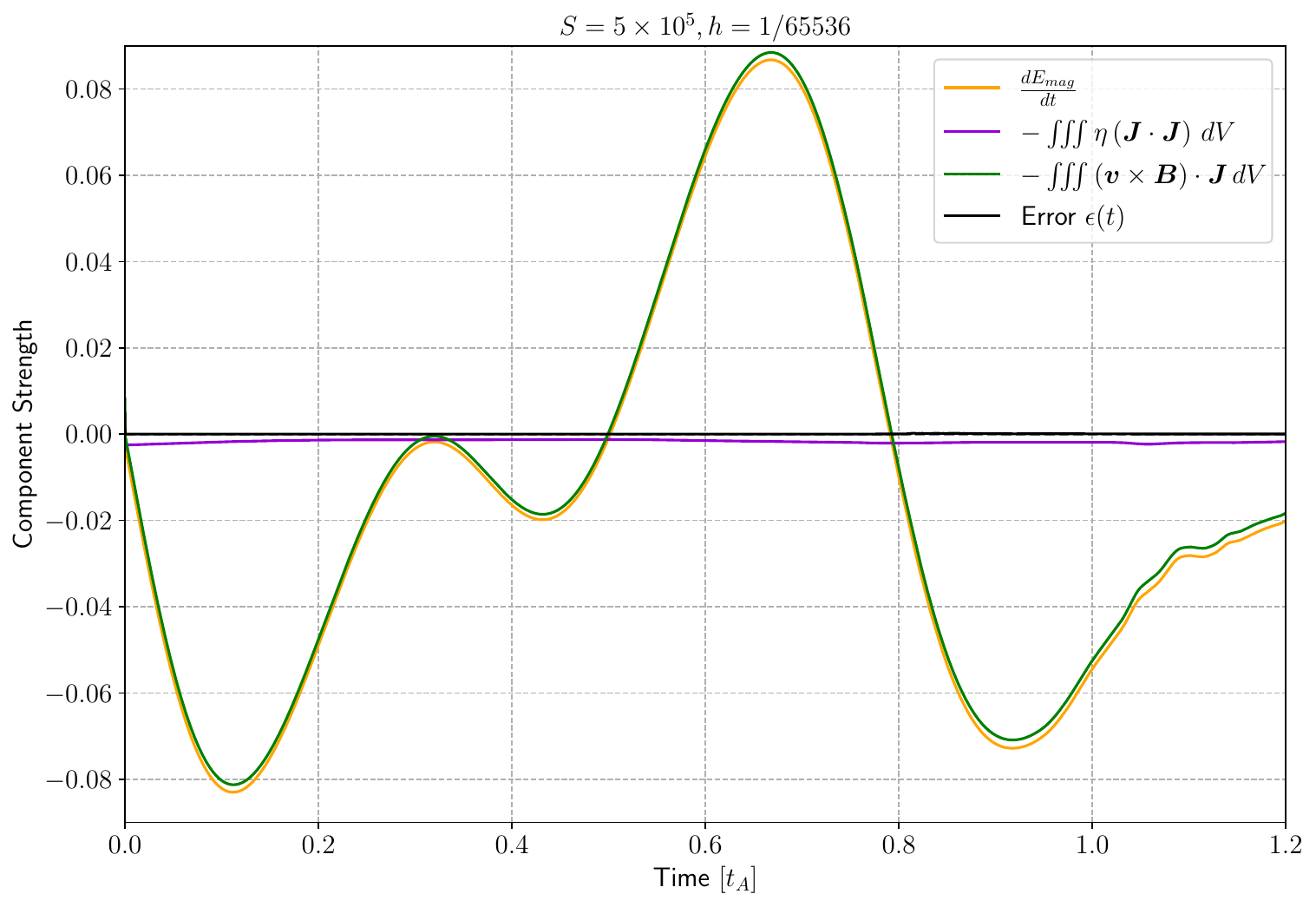}
    \caption{\giovani{Components of the Eq. (\ref{eq:mag_energy_evol}) and numerical error estimation (black solid line) from Eq. (\ref{eq:error_mag_terms}) for the simulation with initial random noise, $S=5 \times 10^5$ and $h=1/65536$.}}
    \label{fig:error_S5e5_h65k}
\end{figure}

\giovani{Finally, in Fig.~\ref{fig:error_S5e5_h65k} we show the evolution of the numerical error and the magnetic terms in Eq.~(\ref{eq:mag_energy_evol}) for the model with $S=5\times 10^{5}$, in which the tearing instability enters the nonlinear regime. In this case, plasmoids start to form at $t \sim 0.5\, t_A$ and merge at $t \sim 0.9\, t_A$ (see Fig.~\ref{fig:bflux_tearing_S3e5_5e5}). As shown in Fig.~\ref{fig:error_S5e5_h65k}, at resolution $h^{-1}=65536$ the numerical error (black solid line) remains smaller in magnitude than all other terms in the magnetic energy density equation. }

For \giovani{the simulations depicted in} Figures \ref{fig:deltaV_S5e4} -- \ref{fig:S1e5_Diff_k_injs}, we not only considered \giovani{the resolutions} satisfying the \citet{Morillo2025} criterion of $\delta/h > 10$, but also analyzed the convergence of the magnetic flux and the numerical error estimated from Eq. (\ref{eq:error_mag_terms}). In these figures, the average reconnection rates were obtained from simulations where the numerical error $\epsilon$ was smaller than the magnitudes of each magnetic term in Eq. (\ref{eq:error_mag_terms}).
\giovani{We note that even when the criterion $\delta/h>10$ is satisfied, plasmoids still form in the high--Lundquist-number regime under controlled numerical error and lead to an enhanced reconnection rate. This finding contrasts with the conclusions reported by \cite{Morillo2025}, who argued that plasmoids are numerical artifacts arising from insufficient resolution.}

\section{Discussion} \label{sec:discussion}

\subsection{Previous Studies of Tearing-Driven Reconnection}

Numerical investigations of magnetic reconnection mediated by the tearing mode have, for decades, been primarily limited to two-dimensional simulations \citep[see, e.g.,][and references therein]{matthaeus1985rapid, biskamp1986magnetic, uzdensky2000two, Samtaney_2009, Daughton_2009_transition, Loureiro2009, bhattacharjee2009fast, huang2010scaling, Loureiro2012, Comisso_2015}. These studies generally indicated that once the Lundquist number exceeds a critical value of $S_c \sim 10^4$, plasmoids emerge, interact, and grow, altering the topology of the current sheet and stabilizing the reconnection rate at a so-called ``universal'' value, $V_{\text{rec}}/V_A \approx S_c^{-1/2} \approx 0.01$.

More recent work, however, has challenged this picture. \citet{Morillo2025}, using high-resolution Orszag--Tang vortex simulations, reported no plasmoids up to $S = 5 \times 10^5$. They argued that when the criterion $\delta/h > 10$ is satisfied---where $\delta$ is the sheet thickness and $h$ the grid cell size---plasmoid formation is suppressed, and the reconnection rate continues to follow the Sweet--Parker scaling, remaining resistivity dependent and therefore slow. On the other hand, the tearing instability is a genuine physical process and should not depend on numerical resolution alone. Indeed, \citet{Huang_2017} demonstrated that in highly controlled setups with extremely low noise amplitudes ($\epsilon = 10^{-30}$), the critical Lundquist number can increase to $S_c \sim 10^6$. This suggests that the absence of plasmoids in \citet{Morillo2025} is more likely a consequence of the extremely quiet numerical environment rather than a genuine suppression of tearing.

These contrasting results highlight the need for simulations that combine high resolution with a controlled level of perturbations and include systematic convergence analyses, which earlier works generally lacked.

\subsection{New Insights into Tearing-Driven Reconnection}

In this work, we performed simulations of high--Lundquist--number current sheets, in which the tearing-mode instability can develop. In such regimes, the sheet thickness decreases as $\delta \sim \eta^{1/2}$ with resistivity, requiring increasingly fine resolution as $\eta$ decreases.\footnote{This setup is relevant to numerical simulations but not directly to astrophysical plasmas, where $\eta$ is fixed by plasma parameters and $S$ grows instead with the macroscopic length scale $L$ (see Sec.~\ref{subsec:applic}).}

To quantify numerical accuracy, we developed a new convergence method based on the full evolution equations of magnetic flux and magnetic energy. By estimating all contributing terms and comparing their sum with the corresponding time derivatives, we directly evaluated the numerical error. This analysis established that the resolution threshold $\delta/h > 10$ is indeed required to resolve the current sheet thickness, consistent with the criterion of \citet{Morillo2025}, but here confirmed through explicit energy-balance tests. Importantly, we introduced random noise into our simulations, thereby ensuring that insufficient perturbations do not artificially suppress tearing. As a result, we clearly observe the onset of tearing instability in well-resolved runs.

Our simulations reveal that plasmoid formation sets in only for \giovani{ $S \gtrsim 2 \times 10^4$.} Beyond this threshold, the reconnection rate is enhanced relative to the Sweet--Parker prediction, $V_{\text{rec}} \sim S^{-1/2}$, but $does$ $not$ saturate to a constant  \giovani{until $S \gtrsim 2 \times 10^5$.
In this range,} it remains Lundquist dependent, following a slower scaling $V_{\text{rec}} \sim S^{-1/3}$, as predicted by linear tearing-mode theory (Eq.~\ref{eq:Vrec_max}) and consistent with analytical expectations \citep[e.g.,][]{lazarian1999reconnection}. \giovani{We have found that fast, resistive-independent reconnection is attained only for $S > 2 \times 10^5$, when then the system enters the nonlinear regime of the tearing-mode instability with the development and merging of plasmoids.  This contrasts with earlier works that predicted this fast, resistive independent regime  for $S \gtrsim 10^4$ \citep[e.g.,][]{bhattacharjee2009fast,Loureiro2012}.}

Thus, even in the plasmoid-unstable regime, our very high-resolution simulations show reconnection rates below the often-cited $0.01 \, V_A$ for %$S > 5 \times 10^4$ 
\giovani{$S \lesssim 2 \times 10^5$}
(Fig. \ref{fig:Average_Vrec_depS-13}). This result \giovani{warns} against interpreting a measured rate of $0.01 \, V_A$ as evidence of convergence, as has sometimes been assumed in the literature. Smaller, Lundquist-dependent rates can persist even at high resolution, indicating that two-dimensional tearing-driven reconnection does not yield a truly fast rate.

\subsection{Limitations of the Present Study}

Our findings establish a new $S^{-1/3}$ regime of two-dimensional linear tearing reconnection that spans a range \giovani{$10^4\lesssim S \lesssim 2 \times 10^5$} of Lundquist numbers, \giovani{followed by a regime of $S$-independent, nonlinear fast tearing-mode reconnection for $S > 2 \times 10^5$}. 
\giovani{These} results show that earlier claims of a universal threshold at $S \approx 10^4$ were not numerically robust, since they were based on insufficiently resolved current sheets.

\giovani{Nevertheless}, our study is limited to resistive MHD in two dimensions. Additional plasma processes---such as anomalous resistivity, pressure anisotropy or kinetic effects---may alter the onset and growth of tearing, either suppressing or enhancing reconnection rates \citep[e.g.,][]{Otto_1991, BirkOtto_1991, Buchner_2006, Huang_2013hyper, ChiouHau_2002, ChiouHau_2003, Ferreira-Santos_2025, Shay_1999, Yamada_2010, Shi_2020, Mirnov_2004, Hosseinpour_2009, Meshcheriakov_2012}. Whether such microphysical effects persist at macroscopic scales remains uncertain. More importantly, reconnection in astrophysical environments is inherently three-dimensional: flux ropes can interact, merge, and drive turbulence \giovani{\citep[see, e.g.,][]{Huang_2016, beg2022evolution, Vicentin_2025}}, leading to reconnection dynamics that cannot be captured in 2D.

Finally, our convergence analysis and numerical error estimation method are themselves restricted in scope. By construction, the method evaluates errors through the resistive MHD magnetic flux and magnetic energy evolution equations. While this provides a rigorous test in 2D and 3D non-ideal single-fluid MHD, its direct applicability to multi-fluid or kinetic simulations is limited, as additional terms and energy channels come into play. Extending such error estimation techniques to more complex plasma descriptions remains an important task for future work.

\subsection{Applicability to Astrophysical Reconnection }
\label{subsec:applic}

The direct applicability of our results to astrophysical environments is limited, since magnetic reconnection in two and three dimensions differs fundamentally \cite[see][]{Lazarian2020review}. In particular, we will show below that the expected Reynolds numbers of the reconnection outflow are so high that the onset of a turbulent regime is inevitable.

In \giovani{our} numerical simulations, the Lundquist number $S$ is increased by decreasing the resistivity $\eta$, while in astrophysical plasmas $\eta$ remains essentially constant and $S$ grows with the macroscopic system size $L$. This distinction has important consequences. Mass conservation requires
\begin{equation}
    \delta \approx L \frac{V_{\rm rec}}{V_A},
\end{equation}
so that the thickness of the outflow depends directly on the reconnection regime. Substituting reconnection scalings, we obtain

\begin{equation}
    V_{\rm rec,SP} \sim V_A S^{-1/2},  \quad \delta_{\rm SP} \sim L S^{-1/2},
\end{equation}

\begin{equation}
    V_{\rm rec,LTM} \sim V_A S^{-1/3}, \quad \delta_{\rm LTM} \sim L S^{-1/3},
\end{equation}

\begin{equation}
    V_{\rm rec, NTM} \simeq 0.01 \, V_A, \quad \delta_{\rm NTM} \sim 0.01 \, L,
\end{equation}

\noindent
corresponding, respectively, to the Sweet--Parker regime (SP, $S \le 10^4$), and the \giovani{linear  ($10^4 < S  \lesssim 2 \times 10^5$) and  nonlinear ($S > 2 \times 10^5$)} tearing-mediated (TM) regimes, the existence of which we confirmed numerically.

The relations above can be expressed in a unified way as
\begin{equation}
    \delta \sim L S^{-\alpha}, \qquad V_{\rm rec} \sim V_A S^{-\alpha},
    \label{eq:scaling}
\end{equation}

\noindent
with $\alpha = 1/2$ for Sweet--Parker, $\alpha = 1/3$ for \giovani{linear} tearing, and $\alpha = 0$ for an $S$-independent regime.

\giovani{In this work, we have assumed a magnetic Prandtl number of $\rm{Pr}_m = \nu /\eta=1$. However, the reconnection rate also depends on $\rm{Pr}_m$. For the Sweet--Parker regime, $V_{\rm rec,SP} \propto \left( 1 + Pr_m \right)^{-1/4}$ \citep{Park_1984_Prm}, while in the tearing-mode reconnection, $V_{\rm rec,TM} \propto \left( 1 + Pr_m \right)^{-1/2}$ \citep{Comisso_2015}} 

From \giovani{relations in Eq.~(\ref{eq:scaling})}, the Reynolds number of the outflow (on the scale of the current sheet thickness $\delta$) is

\begin{equation}
    Re = \frac{V_A \delta}{\nu} \approx \frac{V_{\rm rec}}{V_A} \, Pr_{\rm m}^{-1} S,
\end{equation}

\noindent

\giovani{Substituting the dependence above for $V_{\rm rec}$ on $S$ and $Pr_{\rm m}$, we have, for the Sweet--Parker range:}

\begin{equation}
   \giovani{Re_{\rm SP} \sim \left( 1 + Pr_m \right)^{-1/4} Pr_{\rm m}^{-1} S^{1/2}, \quad S < S_c \approx 10^4,}
\end{equation}

\noindent
for the \giovani{linear} tearing-mediated regime,

\begin{equation}
    \giovani{Re_{\rm LTM} \sim \left( 1 + Pr_m \right)^{-1/2}Pr_{\rm m}^{-1} S^{2/3}, \quad 10^4 < S \lesssim 2 \times 10^5},
\end{equation}

\noindent
and for the \giovani{nonlinear $S-$independent} regime,

\begin{equation}
    \giovani{Re_{\rm NTM} \sim 0.01 \, \left( 1 + Pr_m \right)^{-1/2} Pr_{\rm m}^{-1} S, \quad S > 2 \times 10^5}.
\end{equation}

Thus, for astrophysical plasmas where $S$ can reach $10^{10}$--$10^{20}$, one inevitably finds $Re \gg 1$, ensuring strongly turbulent outflows. \giovani{Moreover, even at the threshold, identified here for the first time, for the onset of nonlinear tearing, $S > 2 \times10^{5}$, the corresponding Reynolds number, $Re\gtrsim 2000$, already indicates a turbulent system.}\footnote{The magnetic Prandtl number $Pr_m$ depends on the medium’s magnetization. Since viscosity in the reconnection outflow is limited by the Bohm diffusion bound, $\nu < r_L v_{th}$ (with $r_L$ the ion Larmor radius and $v_{th}$ the ion thermal speed), typical magnetized astrophysical plasmas have $Pr_m < 1$.} This agrees with arguments by \citet{Lazarian2020review} that reconnection in realistic astrophysical conditions cannot remain laminar (see more details in Lazarian et al., \textit{in prep.}).

The presence of turbulence fundamentally changes reconnection. In 3D, turbulence alone---even without tearing---drives reconnection at a rate independent of resistivity \citep{lazarian1999reconnection, kowal2009numerical, Huang_2016, Beresnyak_2017, kowal2017statistics, Kadowaki_2018, beg2022evolution, Wang_2023, Vicentin_2025}. Such turbulent reconnection relies on field-line wandering induced by Alfvénic modes, which are absent in 2D MHD. Therefore, 2D studies of turbulent effects are not adequate, and any resistivity-independent regime discovered in 2D would not directly apply to astrophysical settings.

Consequently, while our 2D simulations shed new light on tearing-driven reconnection and demonstrate a revised scaling regime, their direct astrophysical applicability is limited. At large Lundquist numbers relevant to astrophysics, reconnection outflows are inevitably turbulent, and only fully three-dimensional studies can capture the interplay between tearing and turbulence that determines \giovani{the reconnection process}.

\giovani{We should also note that, in light of the results obtained in the present work, 
%we can state that 
the enhanced reconnection rates of $V_{\rm rec} / V_A = 0.02-0.04$ reported by \cite{beg2022evolution} and \cite{Vicentin_2025} in the self-generated turbulent regime (SGTR) are likely 
%due to insufficient 
influenced by limited numerical 
resolution in their 3D simulations, since  grid sizes of $h^{-1}=512-2048$ 
%is not enough 
appear insufficient to 
properly  resolve the current sheets at $S=1-2 \times 10^5$.}

\giovani{Assessing the  resolution requirements at such high Lundquist numbers is challenging in 3D MHD simulations, and a uniform grid may be not the most efficient strategy. Therefore, future work is necessary to clarify the role of tearing-mode-driven turbulence in enabling  fast reconnection in fully 3D systems.}

\section{Conclusions}
\label{sec:conclusions}

Despite the large body of previous work devoted to 2D magnetic reconnection in the high-Lundquist-number regime, there has been a notable lack of systematic convergence studies and quantitative estimations of numerical errors. When the Lundquist number $S$ is increased by reducing the resistivity $\eta$, since $S \sim \eta^{-1}$, the current sheet becomes extremely thin, with its initial laminar thickness scaling according to the Sweet--Parker model as $\delta \sim \eta^{1/2}$. This demands extremely high numerical resolution for the current sheet to be accurately resolved. Without careful error estimation and convergence analysis, it is unclear to what extent previously reported reconnection results reflect physical reality or numerical artifacts.  

In this work, we carried out very high-resolution 2D MHD simulations of reconnection layers at Lundquist numbers ranging from \giovani{$10^3$ to $5 \times 10^5$}, where tearing instability can arise. We also introduced a new method to estimate numerical errors directly from the magnetic energy balance equation, allowing us to distinguish between genuine physical behavior and artifacts of discretization.  

Our simulations clearly confirm the theoretical prediction for the onset of plasmoid instability. For $Pr_{\rm m}=1$, we observe the transition from the Sweet--Parker scaling of the reconnection rate to a tearing-dominated regime at \giovani{$S_c \simeq 1.2 \times 10^4$}. This value corresponds to about 5–6 e-foldings of growth of the fastest tearing mode within one Alfvén crossing time, in excellent agreement with the analytic criterion $S_c \approx (N/C_\gamma)^4$ derived in Sec.~\ref{ssec:Critical_Sc}. The quantitative match between simulations and analysis demonstrates the robustness of the growth–before–advection condition as the controlling factor for plasmoid-mediated reconnection.  

Above this threshold, \giovani{for $2 \times 10^4 \lesssim S \lesssim 2 \times 10^5$}, the current sheet fragments into plasmoids and the reconnection rate departs from the Sweet--Parker scaling. Importantly, unlike several earlier reports of an $S$-independent reconnection rate in the plasmoid regime (e.g., \citealt{loureiro2007instability, bhattacharjee2009fast, huang2010scaling}), our converged simulations do \emph{not} exhibit saturation \giovani{for this range of Lundquist number}. Instead, when secondary current sheets are properly resolved and numerical errors are rendered insignificant, the reconnection rate follows the \giovani{linear} tearing-limited prediction (Eq.~\ref{eq:Vrec_max}), 
\begin{equation}
  \frac{V_{\rm rec}}{V_A} \sim S^{-1/3},
\end{equation}
consistent with \citet[][and references therein]{lazarian1999reconnection}.
%and subsequent theoretical expectations. 
This indicates that, within resistive MHD at $Pr_{\rm m} = 1$, \giovani{linear tearing-mediated} reconnection retains a finite dependence on $S$ and therefore remains slow.

\giovani{Nonlinear tearing instability, with plasmoid merging, secondary island formation, and reconnection saturation rate at $V_{\rm rec} \sim 0.01 \, V_A$ is achieved only at $S > 2 \times 10^5$. At these high values, the Reynolds number is large, and turbulence may become important.}

The main results of our numerical study can be summarized as follows:

\begin{itemize}

\item[1.]  
We demonstrate that a resolution threshold of $\delta/h > 10$ (where $h$ is the grid cell size) is necessary to ensure numerical convergence of the magnetic flux. This criterion, originally suggested by \citet{Morillo2025}, is validated and extended in our work through systematic grid refinement, a new quantitative error estimation method based on the magnetic energy density balance (Sec.~\ref{ssec:Error_Estimation}), \giovani{and from the linear theory of tearing-mediated reconnection (Appendix \ref{Appendix:Minimum_Resolution})}. Our analysis shows that only when this convergence requirement is satisfied,  the reconnection rates reflect genuine physics rather than numerical artifacts. This result emphasizes the critical importance of rigorous convergence checks in reconnection studies.

\item[2.]  
The reconnection rate exhibits \giovani{three} distinct regimes as a function of the Lundquist number. For $S \lesssim 10^4$, the reconnection rate follows the classical Sweet--Parker scaling, $V_{\rm rec} \sim S^{-1/2}$, consistent with long-established theoretical predictions \citep{Sweet1958, Parker1957}. For higher values \giovani{of $ 2\times 10^4 \lesssim S \lesssim 2 \times 10^5$}, the tearing instability sets in, and plasmoids are generated. In this regime, the reconnection rate departs from the Sweet--Parker law, but does not saturate to an $S$-independent value. Instead, it follows the tearing-limited scaling $V_{\rm rec} \sim S^{-1/3}$ (Eq.~\ref{eq:Vrec_max}), in agreement with earlier theoretical expectations \citep[][]{lazarian1999reconnection}. \giovani{Nonlinear tearing instability, with plasmoid merging and resistive independent fast reconnection rate with $V_{\rm rec} \sim 0.01 \, V_A$, 
occurs only for $S > 2 \times 10^5$.}

\item[3.]  
For $S \gtrsim 2 \times 10^4$, the tearing instability becomes dynamically important and plasmoids form, even when the simulations satisfy the resolution criterion $\delta/h > 10$. This contradicts the claim by \citet{Morillo2025} that plasmoid formation is suppressed in well-resolved simulations. Our results suggest instead that the absence of plasmoids in those studies is most likely due to insufficient perturbations to trigger the instability, rather than to a fundamental suppression of tearing. This demonstrates that tearing instability is a genuine physical process that cannot be removed simply by increased numerical resolution.

\item[4.] As stressed in Section \ref{subsec:applic}, in realistic astrophysical systems, the Reynolds number of the outflow increases as $Re \sim Pr_{\rm m}^{-1} S^{1-\alpha}$ with $\alpha \in [0, 1/2]$. For high $S$, this implies $Re \gg 1$, ensuring that reconnection occurs in a turbulent regime. Since turbulence fundamentally alters reconnection and behaves differently in 2D and 3D (see \citealt{lazarian1999reconnection, kowal2009numerical, Lazarian2020review}), 2D studies cannot directly clarify the properties of high-$S$ astrophysical reconnection.

\end{itemize}  

Future studies should extend the investigation of tearing-mediated reconnection into the regime of even higher Lundquist numbers, while also moving beyond the 2D approximation to high-resolution 3D simulations. Only then will it be possible to fully address the interplay between plasmoid instability, secondary current sheet formation, and turbulence, and to establish the conditions under which reconnection becomes truly fast in astrophysical environments.

\begin{acknowledgments}

\giovani{The authors acknowledge Luca Comisso, Alexander Russell, Yi-Min Huang, and Diego Falceta-Gonçalves for fruitful discussions.} GHV and EMdGDP acknowledge the support of the Brazilian Funding Agency FAPESP (grants 2013/10559-5, 2020/11891-7, 2021/02120-0, and 2023/10590-1), EMdGDP also acknowledges the support from CNPq (grant 308643/2017-8). GK acknowledges support from FAPESP (grants 2013/10559-5, 2021/02120-0, 2021/06502-4, and 2022/03972-2). AL acknowledges NSF grant AST 2307840. EMdGDP also acknowledges partial support by grant no. NSF PHY-2309135 to the Kavli Institute for Theoretical Physics (KITP) and the fruitful discussions during her stay there. 
The simulations presented in this work were performed using the clusters of the Group of Plasmas and High-Energy Astrophysics at IAG-USP (GAPAE), and of the Group of Theoretical Astrophysics at EACH-USP (Hydra), acquired with support from FAPESP (grants 2013/10559-5, 2021/02120-0, and 2013/04073-2).
\end{acknowledgments}

\newpage

\appendix

\section{Confirmation of the Scaling of the Maximum Growth Rate and Wavenumber}
\label{sec:appendix_growth}

In this Appendix, we present a numerical confirmation of the scaling relations for the maximum growth rate and the corresponding wavenumber of the linear tearing instability. Starting from the incompressible visco-resistive MHD equations, the system can be linearized, yielding the coupled equations for the $z$-components of velocity and magnetic field perturbations \citep{Tenerani_2015}:
\begin{equation}
\label{eq:u}
\gamma \left( \hat{u}'' - k^2 \hat{u} \right) =
 i k \left[ B_0 \left( \hat{b}'' - k^2 \hat{b} \right) - B_0'' \hat{b} \right] + \nu \left( \hat{u}'''' - 2 k^2 \hat{u}'' + k^4 \hat{u} \right)
\end{equation}
\begin{equation}
\label{eq:b}
\gamma \hat{b} = i k B_0 \hat{u} + \eta \left( \hat{b}'' - k^2 \hat{b} \right)
\end{equation}
where $\gamma$ is the growth rate, $k$ is the wavenumber, $\hat{u}$ and $\hat{b}$ are the velocity and magnetic field perturbations, respectively, and $\nu$ and $\eta$ denote viscosity and resistivity. Primes indicate derivatives with respect to $z$ (first, second, and fourth order, respectively). The local Lundquist number is defined as $S_a = V_A a / \eta$, with $\tau_{a} = a/V_A$ the Alfvén time.

The equilibrium is chosen to be the force-free Harris current sheet with magnetic field
\begin{equation}
  \mathbf{B}_0 = B_0(z)\hat{\mathbf{i}} + B_g(z)\hat{\mathbf{j}},
\end{equation}
where $B_0(z) = \tanh(z/a)$ and $B_g(z) = {\rm sech}(z/a)$, such that $\left| \mathbf{B}_0 \right|^2 = 1$. The current sheet half-thickness $a$ is taken as the unit length, and the equilibrium velocity is zero everywhere. This choice ensures that both the reconnecting and guide field components are present, while the total magnetic field strength remains constant.

To solve the eigenvalue problem defined by Eqs.~(\ref{eq:u})--(\ref{eq:b}), we used the \textit{Pseudo-Spectral Eigenvalue Calculator with an Automated Solver} (\texttt{PSECAS}) framework \citep{Berlok_2019}. Perturbations were expanded in normal modes, reducing the problem to a set of ODEs in $z$, discretized with a pseudo-spectral collocation method on a non-uniform grid optimized to resolve the steep gradients near the current sheet center. The eigenvalue problem was solved iteratively, with resolution increased until convergence of growth rates was achieved. The golden section search algorithm was applied to determine the maximum growth rate, with wavenumber relative tolerance set to $10^{-4}$ and growth rate absolute and relative tolerances set to $10^{-8}$ and $10^{-4}$, respectively. \giovani{Additional details on the numerical implementation, grid construction, and convergence tests (including the search for the maximum growth rate) are given in \cite{KowalFalceta:2024} and \cite{Ferreira-Santos_2025}.}

\begin{figure}[h!]
    \centering
    \includegraphics[width=0.99\linewidth]{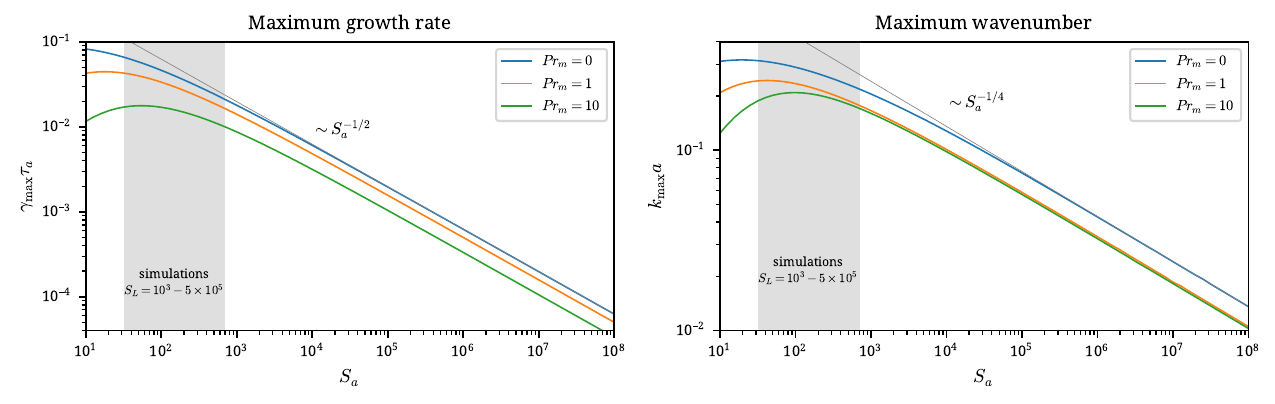}
    \caption{Maximum growth rate (left) and corresponding wavenumber (right) of the tearing instability as a function of Lundquist number $S_a$ in the range $10^{1} \leq S_a \leq 10^{8}$, computed with \texttt{PSECAS} for different Prandtl numbers. The shaded region marks the $S_a$ ($S_a = S^{1/2}$) interval covered in the simulations presented in this work. The grey line shows the analytical scalings, and fitted coefficients $C_\gamma$ and $C_k$ for all Prandtl numbers are reported in the text.}
    \label{fig:appendix_growth}
\end{figure}

Figure~\ref{fig:appendix_growth} shows the maximum growth rate $\gamma_{\max}$ (left panel) and the corresponding wavenumber (right panel) as a function of $S_a$ for several values of the magnetic Prandtl number, $Pr_{\rm m} = \nu/\eta$. The shaded region indicates the range of $S_a$ values covered by our direct numerical simulations in the main part of the paper. From the fits to the high-$S_a$ results we obtained the scaling coefficients $C_\gamma$ and $C_k$ corresponding to the analytical predictions for $\gamma_{\max} \tau_a \simeq C_\gamma S_a^{-1/2}$ and $k_{\max} a \simeq C_k S_a^{-1/4}$. The fitted coefficients are
\begin{align*}
Pr_{\rm m} = 0: &\quad C_\gamma = 0.623,\; C_k = 1.360, \\
Pr_{\rm m} = 1: &\quad C_\gamma = 0.499,\; C_k = 1.053, \\
Pr_{\rm m} = 10: &\quad C_\gamma = 0.333,\; C_k = 1.029.
\end{align*}
For the case $Pr_{\rm m}=1$, used in the main body of this paper, we adopt $C_\gamma \approx 0.5$ and $C_k \approx 1.05$.

It is worth noting that for $S_a < 10^4$, the measured growth rates begin to deviate from the expected asymptotic scaling. This regime includes the Lundquist numbers accessible in our nonlinear numerical simulations, and therefore direct comparison to theory is limited by this departure. Nevertheless, the eigenvalue analysis confirms the validity of the scaling laws in the asymptotic regime and provides consistent estimates of the proportionality constants used in our analysis.

\section{\giovani{Estimation of minimum resolution required to resolve the current sheet }} 
\label{Appendix:Minimum_Resolution}

\begin{figure}[h!]
    \centering
    \includegraphics[width=0.99\linewidth]{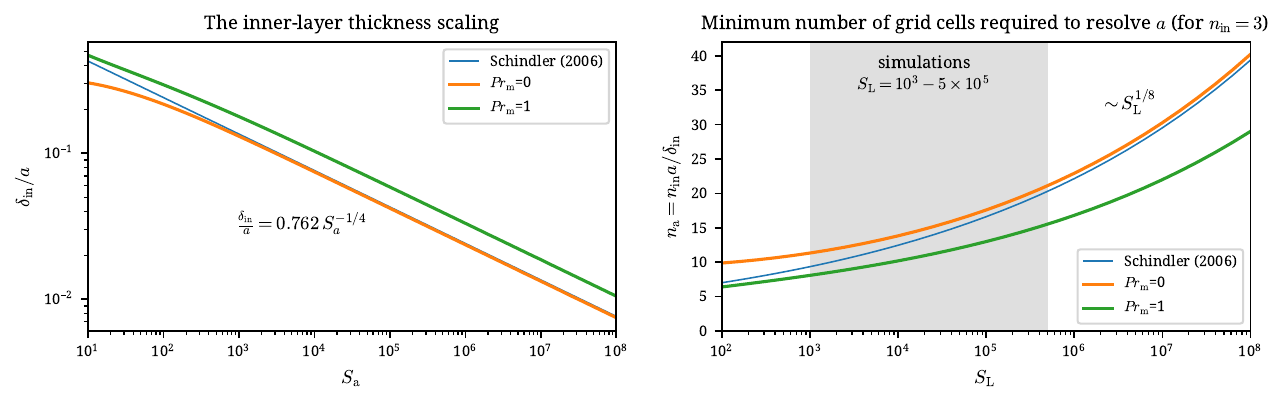}
    \caption{\giovani{Inner-layer thickness and resolution criterion for the fastest-growing tearing mode.  Left: comparison between the theoretical inner-layer thickness $\delta_{\rm in}/a = 0.762 \, S_{\rm a}^{-1/4}$ derived by \cite{Schindler2006} and the numerical estimate obtained from the eigenfunctions of the linear problem, where $\delta_{\rm in}$ is defined as the distance $z_{\rm eq}$ from the resonant surface at which the magnitudes of the ideal and non-ideal terms in the induction equation become equal. Results are shown for Prandtl numbers $Pr_{\rm m}=0$ and $Pr_{\rm m}=1$. The numerical and theoretical predictions exhibit excellent agreement for $S_{\rm a} \gtrsim 10^3$.  Right: required number of grid cells per current-sheet thickness $a$ assuming $n_{\rm in}=3$ cells across $\delta_{\rm in}$, plotted as a function of the large-scale Lundquist number $S_L$ for $Pr_{\rm m}=0$ and $Pr_{\rm m}=1$. Because $\delta_{\rm in} \propto S_L^{-1/8}$, the resolution requirement increases with $S_L$, demonstrating that convergence criteria such as $\delta/h > 10$ remain adequate only at relatively low Lundquist numbers, as the ones considered in this work.}
    }
    \label{fig:appendix_thicknes}
\end{figure}

\giovani{
A practical resolution requirement for resolving the onset of plasmoid-mediated reconnection can be obtained by combining the linear theory of the tearing mode with a numerical measurement of the inner-layer thickness. \cite{Schindler2006} derived that, for the fastest growing mode of the classical resistive tearing instability, the thickness of the inner resistive layer scales as
\begin{equation}
    \frac{\delta_{\rm in}}{a} \simeq 0.762 \, S_a^{-1/4},
    \label{eq:thickness_schindler}
\end{equation}
where $S_a$ is the Lundquist number based on the current-sheet half-thickness $a$. From our numerical eigenproblem solutions, we determine $\delta_{\rm in}$ directly by comparing the ideal term $i k B_0 \hat{u}$ and the non-ideal term $\eta(\hat{b}'' - k^2 \hat{b})$ in the induction equation (Eq.~\ref{eq:b}) as functions of $z$, using the corresponding eigenfunctions of the velocity and magnetic perturbations. We define $z_{\rm eq}$ as the distance from the resonant surface $z=0$ at which the magnitudes of these two terms become equal, and take $\delta_{\rm in} \approx z_{\rm eq}$. The left panel of Figure~\ref{fig:appendix_thicknes} shows that this numerically inferred inner-layer thickness follows the theoretical prediction by Schindler extremely well for $S_a \gtrsim 10^3$ when $Pr_{\rm m}=0$. For $Pr_{\rm m}=1$, the thickness becomes slightly larger, as expected from the effect of viscosity. Once $\delta_{\rm in}$ is known, the minimum grid resolution required to resolve the inner layer follows directly: assuming $n_{\rm in}$ cells across $\delta_{\rm in}$, the number of cells per current-sheet thickness is
\begin{equation}
    n_{\rm a} = n_{\rm in} \frac{a}{\delta_{\rm in}} \approx 1.312 \, n_{\rm in} \, S_a^{1/4} \approx 1.312 \, n_{\rm in} \, S_{\rm L}^{1/8}.
\end{equation}
The right panel of Figure~\ref{fig:appendix_thicknes} shows this requirement as a function of the large-scale Lundquist number $S_L$ for $Pr_{\rm m}=0$ and $Pr_{\rm m}=1$. The commonly adopted criterion $\delta/h > 10$ advocated by \cite{Morillo2025} is therefore sufficient only at relatively modest Lundquist numbers. Because $\delta_{\rm in} \propto S_L^{-1/8}$ (equivalently $\delta_{\rm in} \propto S_a^{-1/4}$), the ratio $a/\delta_{\rm in}$ and thus the required number of cells across the current sheet increases slowly but systematically with $S_L$, implying that substantially higher resolutions are needed to ensure convergence in the high-Lundquist-number regime relevant for plasmoid-dominated reconnection.
}

\section{Long-Term Simulation}
\label{sec:appendix_sim}

\begin{figure*}[ht]
    \centering
    \includegraphics[width = 0.8 \textwidth]{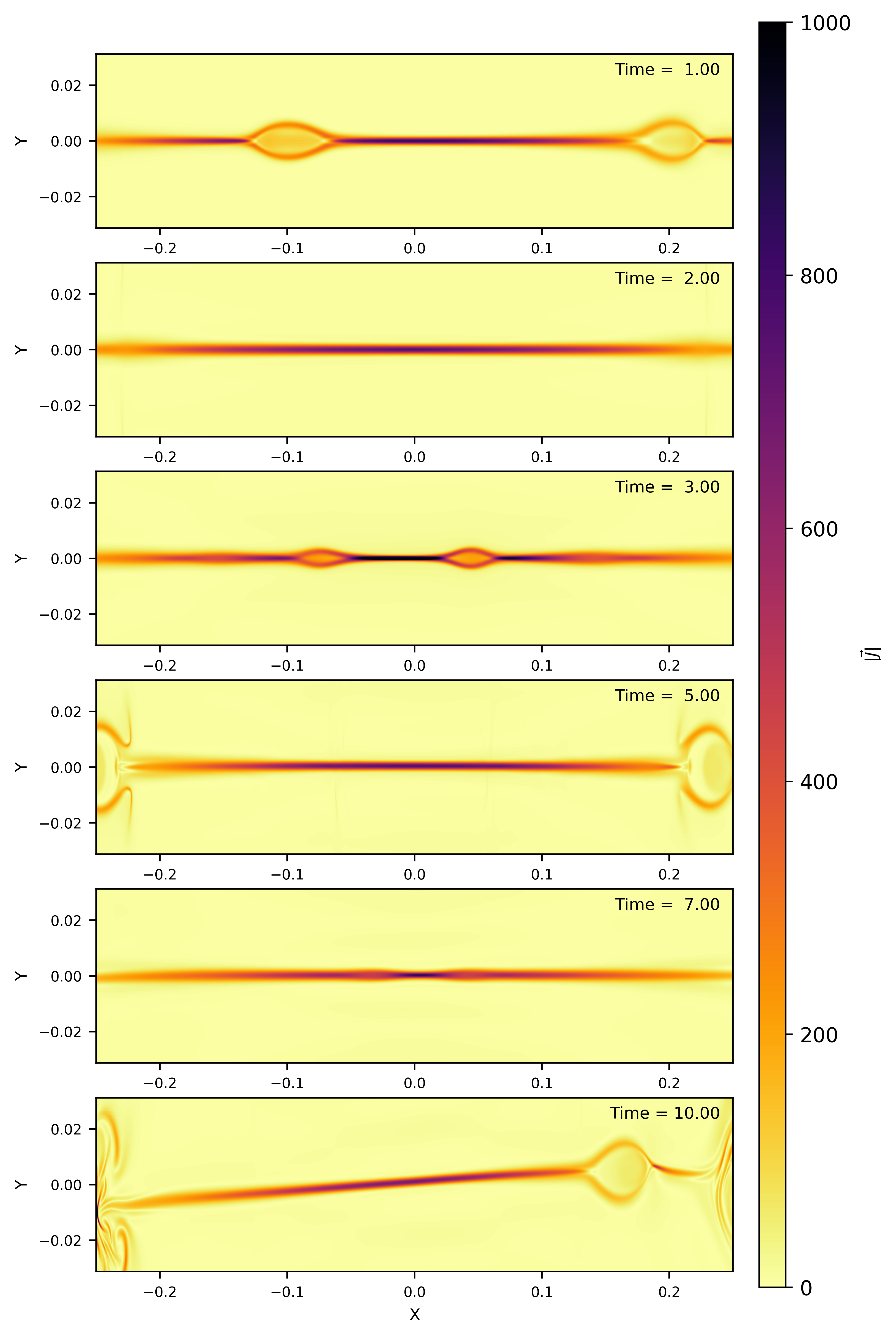}
    \caption{Colormaps of the current density magnitude $|\mathbf{J}| = |\nabla \times \mathbf{B}|$ at different times for the simulation with $S=10^5$ and $h=1/16384$. We zoomed in the region $(x,y) \in [-0.25, 0.25] \times [-0.025, 0.025]$. }
    \label{fig:curdens_S1e5_upt10}
\end{figure*}

%In this appendix, we also show, 
In this appendix, we  show, 
in Fig. \ref{fig:curdens_S1e5_upt10}, the colormaps of the current density magnitude ($|\mathbf{J}| = |\nabla \times \mathbf{B}|$) of our long-term  2D MHD simulation for $S=10^5$ and $h=1/16384$, at different snapshots. We notice that plasmoids are formed in the high-Lundquist number regime, but are advected out of the domain, even in runs lasting up to $t_\text{max} = 10 \, t_A$. \giovani{In this case, the plasmoids do not undergo a merger cascade and do not grow into ``monster'' plasmoids, remaining in the linear phase of tearing-mode instability (see also Fig. \ref{fig:Average_Vrec_depS-13}).}

%\alex{\bf AL. Should we move this into the nain text?}

\bibliography{sample631}{}

@ARTICLE{Giannios2009,
       author = {{Giannios}, Dimitrios and {Uzdensky}, Dmitri A. and {Begelman}, Mitchell C.},
        title = "{Fast TeV variability in blazars: jets in a jet}",
      journal = {\mnras},
         year = 2009,
        month = may,
       volume = {395},
       number = {1},
        pages = {L29-L33},
          doi = {10.1111/j.1745-3933.2009.00635.x},
archivePrefix = {arXiv},
       eprint = {0901.1877},
 primaryClass = {astro-ph.HE},
       adsurl = {https://ui.adsabs.harvard.edu/abs/2009MNRAS.395L..29G}
}

@ARTICLE{zhang-yan2011,
       author = {{Zhang}, Bing and {Yan}, Huirong},
        title = "{The Internal-collision-induced Magnetic Reconnection and Turbulence (ICMART) Model of Gamma-ray Bursts}",
      journal = {\apj},
         year = 2011,
        month = jan,
       volume = {726},
       number = {2},
          eid = {90},
        pages = {90},
          doi = {10.1088/0004-637X/726/2/90},
archivePrefix = {arXiv},
       eprint = {1011.1197},
 primaryClass = {astro-ph.HE},
       adsurl = {https://ui.adsabs.harvard.edu/abs/2011ApJ...726...90Z}
}

@ARTICLE{nalewajko2011MNRAS.413..333N,
       author = {{Nalewajko}, Krzysztof and {Giannios}, Dimitrios and {Begelman}, Mitchell C. and {Uzdensky}, Dmitri A. and {Sikora}, Marek},
        title = "{Radiative properties of reconnection-powered minijets in blazars}",
      journal = {\mnras},
         year = 2011,
        month = may,
       volume = {413},
       number = {1},
        pages = {333-346},
          doi = {10.1111/j.1365-2966.2010.18140.x},
archivePrefix = {arXiv},
       eprint = {1007.3994},
 primaryClass = {astro-ph.HE},
       adsurl = {https://ui.adsabs.harvard.edu/abs/2011MNRAS.413..333N}
}

@ARTICLE{nishkawa2021,
       author = {{Nishikawa}, Kenichi and {Du{\c{t}}an}, Ioana and {K{\"o}hn}, Christoph and {Mizuno}, Yosuke},
        title = "{PIC methods in astrophysics: simulations of relativistic jets and kinetic physics in astrophysical systems}",
      journal = {Living Reviews in Computational Astrophysics},
         year = 2021,
        month = dec,
       volume = {7},
       number = {1},
          eid = {1},
        pages = {1},
          doi = {10.1007/s41115-021-00012-0},
archivePrefix = {arXiv},
       eprint = {2008.02105},
 primaryClass = {astro-ph.HE},
       adsurl = {https://ui.adsabs.harvard.edu/abs/2021LRCA....7....1N}
}

@ARTICLE{dalpino2024,
       author = {{de Gouveia Dal Pino}, Elisabete M. and {Medina-Torrejon}, Tania E.},
        title = "{Particle Acceleration Time due to Turbulent-Induced Magnetic Reconnection}",
      journal = {arXiv e-prints},
         year = 2024,
        month = oct,
          eid = {arXiv:2410.13071},
        pages = {arXiv:2410.13071},
          doi = {10.48550/arXiv.2410.13071},
archivePrefix = {arXiv},
       eprint = {2410.13071},
 primaryClass = {astro-ph.HE},
       adsurl = {https://ui.adsabs.harvard.edu/abs/2024arXiv241013071D}
}

@ARTICLE{kowal2009numerical,
       author = {{Kowal}, Grzegorz and {Lazarian}, A. and {Vishniac}, E.~T. and {Otmianowska-Mazur}, K.},
        title = "{Numerical Tests of Fast Reconnection in Weakly Stochastic Magnetic Fields}",
      journal = {\apj},
         year = 2009,
        month = jul,
       volume = {700},
       number = {1},
        pages = {63-85},
          doi = {10.1088/0004-637X/700/1/63},
archivePrefix = {arXiv},
       eprint = {0903.2052},
 primaryClass = {astro-ph.GA},
       adsurl = {https://ui.adsabs.harvard.edu/abs/2009ApJ...700...63K}
}

@ARTICLE{kowal2012visc,
       author = {{Kowal}, G. and {Lazarian}, A. and {Vishniac}, E.~T. and {Otmianowska-Mazur}, K.},
        title = "{Reconnection studies under different types of turbulence driving}",
      journal = {Nonlinear Processes in Geophysics},
         year = 2012,
        month = apr,
       volume = {19},
       number = {2},
        pages = {297-314},
          doi = {10.5194/npg-19-297-2012},
archivePrefix = {arXiv},
       eprint = {1203.2971},
 primaryClass = {astro-ph.SR},
       adsurl = {https://ui.adsabs.harvard.edu/abs/2012NPGeo..19..297K}
}

@article{Huang_2016,
doi = {10.3847/0004-637X/818/1/20},
url = {https://dx.doi.org/10.3847/0004-637X/818/1/20},
year = {2016},
month = {feb},
publisher = {The American Astronomical Society},
volume = {818},
number = {1},
pages = {20},
author = {Yi-Min Huang and A. Bhattacharjee},
title = {TURBULENT MAGNETOHYDRODYNAMIC RECONNECTION MEDIATED BY THE PLASMOID INSTABILITY},
journal = {The Astrophysical Journal}
}

@article{bhattacharjee2009fast,
  title={Fast reconnection in high-Lundquist-number plasmas due to the plasmoid instability},
  author={Bhattacharjee, A and Huang, Yi-Min and Yang, H and Rogers, B},
  journal={Physics of Plasmas},
  volume={16},
  number={11},
  pages={112102},
  year={2009},
  publisher={American Institute of Physics}
}

@INPROCEEDINGS{Sweet1958,
       author = {{Sweet}, P.~A.},
        title = "{The Neutral Point Theory of Solar Flares}",
    booktitle = {Electromagnetic Phenomena in Cosmical Physics},
         year = 1958,
       editor = {{Lehnert}, B.},
       volume = {6},
        month = jan,
        pages = {123},
       adsurl = {https://ui.adsabs.harvard.edu/abs/1958IAUS....6..123S}
}

@ARTICLE{Parker1957,
       author = {{Parker}, E.~N.},
        title = "{Sweet's Mechanism for Merging Magnetic Fields in Conducting Fluids}",
      journal = {\jgr},
         year = 1957,
        month = dec,
       volume = {62},
       number = {4},
        pages = {509-520},
          doi = {10.1029/JZ062i004p00509},
       adsurl = {https://ui.adsabs.harvard.edu/abs/1957JGR....62..509P}
}

@inproceedings{petschek1964physics,
  title={on the Physics of Solar Flares},
  author={Petschek, HE},
  booktitle={Proc. of AAS-NASA Symp.},
  volume={425},
  year={1964},
  organization={NASA Spec. Pub.}
}

@article{loureiro2007instability,
  title={Instability of current sheets and formation of plasmoid chains},
  author={Loureiro, NF and Schekochihin, AA and Cowley, SC},
  journal={Physics of Plasmas},
  volume={14},
  number={10},
  pages={100703},
  year={2007},
  publisher={American Institute of Physics}
}

@article{lazarian1999reconnection,
  title={Reconnection in a weakly stochastic field},
  author={Lazarian, A and Vishniac, Ethan T},
  journal={The Astrophysical Journal},
  volume={517},
  number={2},
  pages={700},
  year={1999},
  publisher={IOP Publishing}
}

@article{alvelius1999random,
  title={Random forcing of three-dimensional homogeneous turbulence},
  author={Alvelius, Krister},
  journal={Physics of Fluids},
  volume={11},
  number={7},
  pages={1880--1889},
  year={1999},
  publisher={American Institute of Physics}
}

@ARTICLE{kowal2017statistics,
       author = {{Kowal}, Grzegorz and {Falceta-Gon{\c{c}}alves}, Diego A. and {Lazarian}, Alex and {Vishniac}, Ethan T.},
        title = "{Statistics of Reconnection-driven Turbulence}",
      journal = {\apj},
         year = 2017,
        month = apr,
       volume = {838},
       number = {2},
          eid = {91},
        pages = {91},
          doi = {10.3847/1538-4357/aa6001},
archivePrefix = {arXiv},
       eprint = {1611.03914},
 primaryClass = {astro-ph.GA},
       adsurl = {https://ui.adsabs.harvard.edu/abs/2017ApJ...838...91K}
}

@ARTICLE{Lazarian2020review,
       author = {{Lazarian}, Alex and {Eyink}, Gregory L. and {Jafari}, Amir and {Kowal}, Grzegorz and {Li}, Hui and {Xu}, Siyao and {Vishniac}, Ethan T.},
        title = "{3D turbulent reconnection: Theory, tests, and astrophysical implications}",
      journal = {Physics of Plasmas},
         year = 2020,
        month = jan,
       volume = {27},
       number = {1},
          eid = {012305},
        pages = {012305},
          doi = {10.1063/1.5110603},
archivePrefix = {arXiv},
       eprint = {2001.00868},
 primaryClass = {astro-ph.HE},
       adsurl = {https://ui.adsabs.harvard.edu/abs/2020PhPl...27a2305L}
}

@article{furth1963finite,
    author = {Furth, Harold P. and Killeen, John and Rosenbluth, Marshall N.},
    title = {Finite‐Resistivity Instabilities of a Sheet Pinch},
    journal = {The Physics of Fluids},
    volume = {6},
    number = {4},
    pages = {459-484},
    year = {1963},
    month = {04},
    issn = {0031-9171},
    doi = {10.1063/1.1706761},
    url = {https://doi.org/10.1063/1.1706761},
    eprint = {https://pubs.aip.org/aip/pfl/article-pdf/6/4/459/12485401/459\_1\_online.pdf},
}

@article{Beresnyak_2017,
doi = {10.3847/1538-4357/834/1/47},
url = {https://dx.doi.org/10.3847/1538-4357/834/1/47},
year = {2016},
month = {dec},
publisher = {The American Astronomical Society},
volume = {834},
number = {1},
pages = {47},
author = {Andrey Beresnyak},
title = {THREE-DIMENSIONAL SPONTANEOUS MAGNETIC RECONNECTION},
journal = {The Astrophysical Journal}
}

@article{Kadowaki_2018,
doi = {10.3847/1538-4357/aad4ff},
url = {https://dx.doi.org/10.3847/1538-4357/aad4ff},
year = {2018},
month = {aug},
publisher = {The American Astronomical Society},
volume = {864},
number = {1},
pages = {52},
author = {L. H. S. Kadowaki and E. M. de Gouveia Dal Pino and J. M. Stone},
title = {{MHD Instabilities in Accretion Disks and Their Implications in Driving Fast Magnetic Reconnection}},
journal = {The Astrophysical Journal}
}

@article{Kadowaki_2021,
doi = {10.3847/1538-4357/abee7a},
url = {https://dx.doi.org/10.3847/1538-4357/abee7a},
year = {2021},
month = {may},
publisher = {The American Astronomical Society},
volume = {912},
number = {2},
pages = {109},
author = {L. H. S. Kadowaki and E. M. de Gouveia Dal Pino and T. E. Medina-Torrejón and Y. Mizuno and P. Kushwaha},
title = {Fast Magnetic Reconnection Structures in Poynting Flux-dominated Jets},
journal = {The Astrophysical Journal}
}

@ARTICLE{Medina-Torrejón_2021,
       author = {{Medina-Torrej{\'o}n}, Tania E. and {de Gouveia Dal Pino}, Elisabete M. and {Kadowaki}, Luis H.~S. and {Kowal}, Grzegorz and {Singh}, Chandra B. and {Mizuno}, Yosuke},
        title = "{Particle Acceleration by Relativistic Magnetic Reconnection Driven by Kink Instability Turbulence in Poynting Flux-Dominated Jets}",
      journal = {\apj},
         year = 2021,
        month = feb,
       volume = {908},
       number = {2},
          eid = {193},
        pages = {193},
          doi = {10.3847/1538-4357/abd6c2},
archivePrefix = {arXiv},
       eprint = {2009.08516},
 primaryClass = {astro-ph.HE},
       adsurl = {https://ui.adsabs.harvard.edu/abs/2021ApJ...908..193M}
}

@ARTICLE{Medina-Torrejón_2023,
       author = {{Medina-Torrej{\'o}n}, Tania E. and {de Gouveia Dal Pino}, Elisabete M. and {Kowal}, Grzegorz},
        title = "{Particle Acceleration by Magnetic Reconnection in Relativistic Jets: The Transition from Small to Large Scales}",
      journal = {\apj},
         year = 2023,
        month = aug,
       volume = {952},
       number = {2},
          eid = {168},
        pages = {168},
          doi = {10.3847/1538-4357/acd699},
archivePrefix = {arXiv},
       eprint = {2303.08780},
 primaryClass = {astro-ph.HE},
       adsurl = {https://ui.adsabs.harvard.edu/abs/2023ApJ...952..168M}
}

@article{parker1988nanoflares,
  title={Nanoflares and the solar X-ray corona},
  author={Parker, Eugene N},
  journal={Astrophysical Journal},
  volume={330},
  pages={474--479},
  year={1988}
}

@article{heyvaerts1984coronal,
  title={Coronal heating by reconnection in DC current systems-A theory based on Taylor's hypothesis},
  author={Heyvaerts, J and Priest, ER},
  journal={Astronomy and Astrophysics},
  volume={137},
  pages={63--78},
  year={1984}
}

@article{lee1985theory,
  title={A theory of magnetic flux transfer at the Earth's magnetopause},
  author={Lee, LC and Fu, ZF},
  journal={Geophysical Research Letters},
  volume={12},
  number={2},
  pages={105--108},
  year={1985},
  publisher={Wiley Online Library}
}

@article{lee1986multiple,
  title={Multiple X line reconnection: 1. A criterion for the transition from a single X line to a multiple X line reconnection},
  author={Lee, LC and Fu, ZF},
  journal={Journal of Geophysical Research: Space Physics},
  volume={91},
  number={A6},
  pages={6807--6815},
  year={1986},
  publisher={Wiley Online Library}
}

@article{beg2022evolution,
  title={Evolution, structure, and topology of self-generated turbulent reconnection layers},
  author={Beg, Raheem and Russell, Alexander JB and Hornig, Gunnar},
  journal={The Astrophysical Journal},
  volume={940},
  number={1},
  pages={94},
  year={2022},
  publisher={IOP Publishing}
}

@article{huang2010scaling,
  title={Scaling laws of resistive magnetohydrodynamic reconnection in the high-Lundquist-number, plasmoid-unstable regime},
  author={Huang, Yi-Min and Bhattacharjee, A},
  journal={Physics of Plasmas},
  volume={17},
  number={6},
  year={2010},
  url = {https://doi.org/10.1063/1.3420208},
  publisher={AIP Publishing}
}

@article{Comisso_2015,
    author = {Comisso, L. and Grasso, D. and Waelbroeck, F. L.},
    title = "{Extended theory of the Taylor problem in the plasmoid-unstable regime}",
    journal = {Physics of Plasmas},
    volume = {22},
    number = {4},
    pages = {042109},
    year = {2015},
    month = {04},
    issn = {1070-664X},
    doi = {10.1063/1.4918331},
    url = {https://doi.org/10.1063/1.4918331}
}

@article{Samtaney_2009,
  title = {Formation of Plasmoid Chains in Magnetic Reconnection},
  author = {Samtaney, R. and Loureiro, N. F. and Uzdensky, D. A. and Schekochihin, A. A. and Cowley, S. C.},
  journal = {Phys. Rev. Lett.},
  volume = {103},
  issue = {10},
  pages = {105004},
  numpages = {4},
  year = {2009},
  month = {Sep},
  publisher = {American Physical Society},
  doi = {10.1103/PhysRevLett.103.105004},
  url = {https://link.aps.org/doi/10.1103/PhysRevLett.103.105004}
}

@article{biskamp1986magnetic,
  title={Magnetic reconnection via current sheets},
  author={Biskamp, Dieter},
  journal={The Physics of fluids},
  volume={29},
  number={5},
  pages={1520--1531},
  year={1986},
  publisher={American Institute of Physics}
}

@article{uzdensky2000two,
  title={Two-dimensional numerical simulation of the resistive reconnection layer},
  author={Uzdensky, DA and Kulsrud, RM},
  journal={Physics of Plasmas},
  volume={7},
  number={10},
  pages={4018--4030},
  year={2000},
  publisher={AIP Publishing}
}

@article{matthaeus1985rapid,
  title={Rapid magnetic reconnection caused by finite amplitude fluctuations},
  author={Matthaeus, WH and Lamkin, SL},
  journal={The Physics of fluids},
  volume={28},
  number={1},
  pages={303--307},
  year={1985},
  publisher={AIP Publishing}
}

@article{Daughton_2009_transition,
  title = {Transition from collisional to kinetic regimes in large-scale reconnection layers},
  author = {Daughton, W. and Roytershteyn, V. and Albright, B. J. and Karimabadi, H. and Yin, L. and Bowers, Kevin J.},
  journal = {Phys. Rev. Lett.},
  volume = {103},
  issue = {6},
  pages = {065004},
  numpages = {4},
  year = {2009},
  month = {Aug},
  publisher = {American Physical Society},
  doi = {10.1103/PhysRevLett.103.065004},
  url = {https://link.aps.org/doi/10.1103/PhysRevLett.103.065004}
}

@ARTICLE{Loureiro2012,
       author = {{Loureiro}, N.~F. and {Samtaney}, R. and {Schekochihin}, A.~A. and {Uzdensky}, D.~A.},
        title = "{Magnetic reconnection and stochastic plasmoid chains in high-Lundquist-number plasmas}",
      journal = {Physics of Plasmas},
         year = 2012,
        month = apr,
       volume = {19},
       number = {4},
        pages = {042303-042303},
          doi = {10.1063/1.3703318},
archivePrefix = {arXiv},
       eprint = {1108.4040},
 primaryClass = {astro-ph.SR},
       adsurl = {https://ui.adsabs.harvard.edu/abs/2012PhPl...19d2303L}
}

@article{Loureiro2009,
    author = {Loureiro, N. F. and Uzdensky, D. A. and Schekochihin, A. A. and Cowley, S. C. and Yousef, T. A.},
    title = "{Turbulent magnetic reconnection in two dimensions}",
    journal = {Monthly Notices of the Royal Astronomical Society: Letters},
    volume = {399},
    number = {1},
    pages = {L146-L150},
    year = {2009},
    month = {10},
    issn = {1745-3925},
    doi = {10.1111/j.1745-3933.2009.00742.x},
    url = {https://doi.org/10.1111/j.1745-3933.2009.00742.x},
    eprint = {https://academic.oup.com/mnrasl/article-pdf/399/1/L146/54677529/mnrasl\_399\_1\_l146.pdf},
}

@ARTICLE{mp5,
       author = {{Suresh}, A. and {Huynh}, H.~T.},
        title = "{Accurate Monotonicity-Preserving Schemes with Runge Kutta Time Stepping}",
      journal = {Journal of Computational Physics},
         year = 1997,
        month = sep,
       volume = {136},
       number = {1},
        pages = {83-99},
          doi = {10.1006/jcph.1997.5745},
       adsurl = {https://ui.adsabs.harvard.edu/abs/1997JCoPh.136...83S}
}

@ARTICLE{SSPRK324,
       author = {{Ranocha}, Hendrik and {Dalcin}, Lisandro and {Parsani}, Matteo and {Ketcheson}, David I.},
        title = "{Optimized Runge-Kutta Methods with Automatic Step Size Control for Compressible Computational Fluid Dynamics}",
      journal = {Communications on Applied Mathematics and Computation},
         year = 2022,
       volume = {4},
        pages = {2661-8893},
          doi = {10.1007/s42967-021-00159-w},
       adsurl = {https://ui.adsabs.harvard.edu/abs/2021arXiv210406836R},
}

@article{Wang_2023,
doi = {10.3847/2041-8213/acf19d},
url = {https://dx.doi.org/10.3847/2041-8213/acf19d},
year = {2023},
month = {sep},
publisher = {The American Astronomical Society},
volume = {954},
number = {2},
pages = {L36},
author = {Wang, Yulei and Cheng, Xin and Ding, Mingde and Liu, Zhaoyuan and Liu, Jian and Zhu, Xiaojue},
title = {Three-dimensional Turbulent Reconnection within the Solar Flare Current Sheet},
journal = {The Astrophysical Journal Letters}
}

@ARTICLE{Morillo2025,
       author = {{Morillo}, Jos{\'e} Mar{\'\i}a Garc{\'\i}a and {Alexakis}, Alexandros},
        title = "{Magnetic reconnection, plasmoids and numerical resolution}",
      journal = {Journal of Fluid Mechanics},
         year = 2025,
        month = mar,
       volume = {1007},
          eid = {R3},
        pages = {R3},
          doi = {10.1017/jfm.2025.109},
archivePrefix = {arXiv},
       eprint = {2406.08951},
 primaryClass = {physics.plasm-ph},
       adsurl = {https://ui.adsabs.harvard.edu/abs/2025JFM..1007R...3M}
}

@ARTICLE{Kulpa-Dybel-2010,
       author = {{Kulpa-Dybe{\l}}, K. and {Kowal}, G. and {Otmianowska-Mazur}, K. and {Lazarian}, A. and {Vishniac}, E.},
        title = "{Reconnection in weakly stochastic B-fields in 2D}",
      journal = {\aap},
         year = 2010,
        month = may,
       volume = {514},
          eid = {A26},
        pages = {A26},
          doi = {10.1051/0004-6361/200913218},
archivePrefix = {arXiv},
       eprint = {0909.1265},
 primaryClass = {astro-ph.GA},
       adsurl = {https://ui.adsabs.harvard.edu/abs/2010A&A...514A..26K}
}

@ARTICLE{PriestForbes2002,
       author = {{Priest}, E.~R. and {Forbes}, T.~G.},
        title = "{The magnetic nature of solar flares}",
      journal = {\aapr},
         year = 2002,
        month = jan,
       volume = {10},
       number = {4},
        pages = {313-377},
          doi = {10.1007/s001590100013},
       adsurl = {https://ui.adsabs.harvard.edu/abs/2002A&ARv..10..313P}
}

@article{Vicentin_2025,
doi = {10.3847/1538-4357/addc62},
url = {https://dx.doi.org/10.3847/1538-4357/addc62},
year = {2025},
month = {jul},
publisher = {The American Astronomical Society},
volume = {987},
number = {2},
pages = {213},
author = {Vicentin, Giovani H. and Kowal, Grzegorz and de Gouveia Dal Pino, Elisabete M. and Lazarian, Alex},
title = {Investigating Turbulence Effects on Magnetic Reconnection Rates through 3D Resistive Magnetohydrodynamic Simulations},
journal = {The Astrophysical Journal}
}

@ARTICLE{Tenerani_2015,
       author = {{Tenerani}, Anna and {Rappazzo}, Antonio Franco and {Velli}, Marco and {Pucci}, Fulvia},
        title = "{The Tearing Mode Instability of Thin Current Sheets: the Transition to Fast Reconnection in the Presence of Viscosity}",
      journal = {\apj},
         year = 2015,
        month = mar,
       volume = {801},
       number = {2},
          eid = {145},
        pages = {145},
          doi = {10.1088/0004-637X/801/2/145},
archivePrefix = {arXiv},
       eprint = {1412.0047},
 primaryClass = {physics.plasm-ph},
       adsurl = {https://ui.adsabs.harvard.edu/abs/2015ApJ...801..145T}
}

@ARTICLE{Yamada1997,
       author = {{Yamada}, Masaaki and {Ji}, Hantao and {Hsu}, Scott and {Carter}, Troy and {Kulsrud}, Russell and {Bretz}, Norton and {Jobes}, Forrest and {Ono}, Yasushi and {Perkins}, Francis},
        title = "{Study of driven magnetic reconnection in a laboratory plasma}",
      journal = {Physics of Plasmas},
         year = 1997,
        month = may,
       volume = {4},
       number = {5},
        pages = {1936-1944},
          doi = {10.1063/1.872336},
       adsurl = {https://ui.adsabs.harvard.edu/abs/1997PhPl....4.1936Y}
}

@ARTICLE{Yamada1994,
       author = {{Yamada}, M. and {Levinton}, F.~M. and {Pomphrey}, N. and {Budny}, R. and {Manickam}, J. and {Nagayama}, Y.},
        title = "{Investigation of magnetic reconnection during a sawtooth crash in a high-temperature tokamak plasma}",
      journal = {Physics of Plasmas},
         year = 1994,
        month = oct,
       volume = {1},
       number = {10},
        pages = {3269-3276},
          doi = {10.1063/1.870479},
       adsurl = {https://ui.adsabs.harvard.edu/abs/1994PhPl....1.3269Y}
}

@ARTICLE{Masuda1994Nature,
       author = {{Masuda}, S. and {Kosugi}, T. and {Hara}, H. and {Tsuneta}, S. and {Ogawara}, Y.},
        title = "{A loop-top hard X-ray source in a compact solar flare as evidence for magnetic reconnection}",
      journal = {\nat},
         year = 1994,
        month = oct,
       volume = {371},
       number = {6497},
        pages = {495-497},
          doi = {10.1038/371495a0},
       adsurl = {https://ui.adsabs.harvard.edu/abs/1994Natur.371..495M}
}

@article{Taylor1986,
  title = {Relaxation and magnetic reconnection in plasmas},
  author = {Taylor, J. B.},
  journal = {Rev. Mod. Phys.},
  volume = {58},
  issue = {3},
  pages = {741--763},
  numpages = {0},
  year = {1986},
  month = {Jul},
  publisher = {American Physical Society},
  doi = {10.1103/RevModPhys.58.741},
  url = {https://link.aps.org/doi/10.1103/RevModPhys.58.741}
}

@article{Uzdensky_2010,
  title = {Fast Magnetic Reconnection in the Plasmoid-Dominated Regime},
  author = {Uzdensky, D. A. and Loureiro, N. F. and Schekochihin, A. A.},
  journal = {Phys. Rev. Lett.},
  volume = {105},
  issue = {23},
  pages = {235002},
  numpages = {4},
  year = {2010},
  month = {Dec},
  publisher = {American Physical Society},
  doi = {10.1103/PhysRevLett.105.235002},
  url = {https://link.aps.org/doi/10.1103/PhysRevLett.105.235002}
}

@ARTICLE{Orszag_Tang_1979,
       author = {{Orszag}, S.~A. and {Tang}, C. -M.},
        title = "{Small-scale structure of two-dimensional magnetohydrodynamic turbulence}",
      journal = {Journal of Fluid Mechanics},
         year = 1979,
        month = jan,
       volume = {90},
        pages = {129-143},
          doi = {10.1017/S002211207900210X},
       adsurl = {https://ui.adsabs.harvard.edu/abs/1979JFM....90..129O}
}

@article{Jara_Almonte_2016,
  title = {Laboratory Observation of Resistive Electron Tearing in a Two-Fluid Reconnecting Current Sheet},
  author = {Jara-Almonte, Jonathan and Ji, Hantao and Yamada, Masaaki and Yoo, Jongsoo and Fox, William},
  journal = {Phys. Rev. Lett.},
  volume = {117},
  issue = {9},
  pages = {095001},
  numpages = {5},
  year = {2016},
  month = {Aug},
  publisher = {American Physical Society},
  doi = {10.1103/PhysRevLett.117.095001},
  url = {https://link.aps.org/doi/10.1103/PhysRevLett.117.095001}
}

@article{Huang_2017,
doi = {10.3847/1538-4357/aa906d},
url = {https://dx.doi.org/10.3847/1538-4357/aa906d},
year = {2017},
month = {nov},
publisher = {The American Astronomical Society},
volume = {849},
number = {2},
pages = {75},
author = {Huang, Yi-Min and Comisso, Luca and Bhattacharjee, A.},
title = {Plasmoid Instability in Evolving Current Sheets and Onset of Fast Reconnection},
journal = {The Astrophysical Journal}
}

@book{kivelson1995introduction,
  title={Introduction to space physics},
  author={Kivelson, Margaret G and Russell, Christopher T},
  year={1995},
  publisher={Cambridge university press}
}

@ARTICLE{Coppi1976,
       author = {{Coppi}, B. and {Galvao}, R. and {Pellat}, R. and {Rosenbluth}, M. and {Rutherford}, P.},
        title = "{Resistive internal kink modes}",
      journal = {Soviet Journal of Plasma Physics},
         year = 1976,
        month = nov,
       volume = {2},
        pages = {533-535},
       adsurl = {https://ui.adsabs.harvard.edu/abs/1976SvJPP...2..533C}
}

@article{dGDP2010,
	author = {de Gouveia Dal Pino, E. M. and Piovezan, P. P. and Kadowaki, L. H. S.},
	title = {The role of magnetic reconnection on jet/accretion disk systems},
	DOI= "10.1051/0004-6361/200913462",
	url= "https://doi.org/10.1051/0004-6361/200913462",
	journal = {A\&A},
	year = {2010a},
	volume = 518,
	pages = "A5",
	month = "",
}

@article{dGDP_Lazarian_2005,
	author = {de Gouveia Dal Pino, E. M. and Lazarian, A.},
	title = {Production of the large scale superluminal ejections  of the
microquasar GRS 1915+105   by violent magnetic reconnection },
	DOI= "10.1051/0004-6361:20042590",
	url= "https://doi.org/10.1051/0004-6361:20042590",
	journal = {A\&A},
	year = 2005,
	volume = 441,
	number = 3,
	pages = "845-853",
}

@article{Kadowaki_2015,
doi = {10.1088/0004-637X/802/2/113},
url = {https://dx.doi.org/10.1088/0004-637X/802/2/113},
year = {2015},
month = {mar},
publisher = {The American Astronomical Society},
volume = {802},
number = {2},
pages = {113},
author = {Kadowaki, L. H. S. and Pino, E. M. de Gouveia Dal and Singh, C. B.},
title = {THE ROLE OF FAST MAGNETIC RECONNECTION ON THE RADIO AND GAMMA-RAY EMISSION FROM THE NUCLEAR REGIONS OF MICROQUASARS AND LOW LUMINOSITY AGNs},
journal = {The Astrophysical Journal}
}

@article{Giannios_2010,
    author = {Giannios, Dimitrios},
    title = {UHECRs from magnetic reconnection in relativistic jets},
    journal = {Monthly Notices of the Royal Astronomical Society: Letters},
    volume = {408},
    number = {1},
    pages = {L46-L50},
    year = {2010},
    month = {10},
    issn = {1745-3925},
    doi = {10.1111/j.1745-3933.2010.00925.x},
    url = {https://doi.org/10.1111/j.1745-3933.2010.00925.x},
    eprint = {https://academic.oup.com/mnrasl/article-pdf/408/1/L46/54672079/mnrasl\_408\_1\_l46.pdf},
}

@article{Khiali_2015,
    author = {Khiali, B. and de Gouveia Dal Pino, E. M. and del Valle, M. V.},
    title = {A magnetic reconnection model for explaining the multiwavelength emission of the microquasars Cyg X-1 and Cyg X-3},
    journal = {Monthly Notices of the Royal Astronomical Society},
    volume = {449},
    number = {1},
    pages = {34-48},
    year = {2015},
    month = {03},
    issn = {0035-8711},
    doi = {10.1093/mnras/stv248},
    url = {https://doi.org/10.1093/mnras/stv248},
    eprint = {https://academic.oup.com/mnras/article-pdf/449/1/34/4130786/stv248.pdf},
}

@article{Cerutti_2012,
doi = {10.1088/0004-637X/746/2/148},
url = {https://dx.doi.org/10.1088/0004-637X/746/2/148},
year = {2012},
month = {feb},
publisher = {The American Astronomical Society},
volume = {746},
number = {2},
pages = {148},
author = {Cerutti, Benoît and Uzdensky, Dmitri A. and Begelman, Mitchell C.},
title = {EXTREME PARTICLE ACCELERATION IN MAGNETIC RECONNECTION LAYERS: APPLICATION TO THE GAMMA-RAY FLARES IN THE CRAB NEBULA},
journal = {The Astrophysical Journal}
}

@article{dGDP_2010b,
author = {de Gouveia Dal Pino, E. M. and Kowal, G. and Kadowaki, L. H. S. and Piovezan, P. and Lazarian, A.},
title = {MAGNETIC FIELD EFFECTS NEAR THE LAUNCHING REGION OF ASTROPHYSICAL JETS},
journal = {International Journal of Modern Physics D},
volume = {19},
number = {06},
pages = {729-739},
year = {2010b},
doi = {10.1142/S0218271810016920},
URL = {https://doi.org/10.1142/S0218271810016920},
eprint = {https://doi.org/10.1142/S0218271810016920}
}

@ARTICLE{porcelli1987viscous,
       author = {{Porcelli}, Francesco},
        title = "{Viscous resistive magnetic reconnection}",
      journal = {Physics of Fluids},
         year = 1987,
        month = jun,
       volume = {30},
       number = {6},
        pages = {1734-1742},
          doi = {10.1063/1.866240},
       adsurl = {https://ui.adsabs.harvard.edu/abs/1987PhFl...30.1734P}
}

@ARTICLE{Lipps1963,
       author = {{Lipps}, Frank B.},
        title = "{Stability of Jets in a Divergent Barotropic Fluid.}",
      journal = {Journal of the Atmospheric Sciences},
         year = 1963,
        month = mar,
       volume = {20},
       number = {2},
        pages = {120-129},
          doi = {10.1175/1520-0469(1963)020<0120:SOJIAD>2.0.CO;2},
       adsurl = {https://ui.adsabs.harvard.edu/abs/1963JAtS...20..120L}
}

@ARTICLE{Goldsmith1970,
       author = {{Goldsmith}, D.~W.},
        title = "{Thermal Instabilities in Interstellar Gas Heated by Cosmic Rays}",
      journal = {\apj},
         year = 1970,
        month = jul,
       volume = {161},
        pages = {41},
          doi = {10.1086/150511},
       adsurl = {https://ui.adsabs.harvard.edu/abs/1970ApJ...161...41G}
}

@ARTICLE{Syrovatskii_1981,
       author = {{Syrovatskii}, S.~I.},
        title = "{Pinch sheets and reconnection in astrophysics}",
      journal = {\araa},
         year = 1981,
        month = jan,
       volume = {19},
        pages = {163-229},
          doi = {10.1146/annurev.aa.19.090181.001115},
       adsurl = {https://ui.adsabs.harvard.edu/abs/1981ARA&A..19..163S}
}

@ARTICLE{Berlok_2019,
       author = {{Berlok}, Thomas and {Pfrommer}, Christoph},
        title = "{On the Kelvin-Helmholtz instability with smooth initial conditions - linear theory and simulations}",
      journal = {Monthly Notices of the Royal Astronomical Society},
         year = 2019,
        month = may,
       volume = {485},
       number = {1},
        pages = {908-923},
          doi = {10.1093/mnras/stz379},
archivePrefix = {arXiv},
       eprint = {1902.01403},
 primaryClass = {astro-ph.GA},
       adsurl = {https://ui.adsabs.harvard.edu/abs/2019MNRAS.485..908B}
}

@ARTICLE{Ferreira-Santos_2025,
       author = {{Ferreira-Santos}, Gabriel L. and {Kowal}, Grzegorz and {Falceta-Gon{\c{c}}alves}, Diego A.},
        title = "{Unveiling a New $β$-Scaling of the Tearing Instability in Weakly Collisional Plasmas}",
      journal = {arXiv e-prints},
         year = 2025,
        month = mar,
          eid = {arXiv:2503.12702},
        pages = {arXiv:2503.12702},
          doi = {10.48550/arXiv.2503.12702},
archivePrefix = {arXiv},
       eprint = {2503.12702},
 primaryClass = {physics.plasm-ph},
       adsurl = {https://ui.adsabs.harvard.edu/abs/2025arXiv250312702F}
}

@ARTICLE{Otto_1991,
       author = {{Otto}, A.},
        title = "{The resistive tearing instability for generalized resistivity models: Theory}",
      journal = {Physics of Fluids B},
         year = 1991,
        month = jul,
       volume = {3},
       number = {7},
        pages = {1739-1745},
          doi = {10.1063/1.859692},
       adsurl = {https://ui.adsabs.harvard.edu/abs/1991PhFlB...3.1739O}
}

@ARTICLE{BirkOtto_1991,
       author = {{Birk}, G.~T. and {Otto}, A.},
        title = "{The resistive tearing instability for generalized resistivity models: Applications}",
      journal = {Physics of Fluids B},
         year = 1991,
        month = jul,
       volume = {3},
       number = {7},
        pages = {1746-1754},
          doi = {10.1063/1.859693},
       adsurl = {https://ui.adsabs.harvard.edu/abs/1991PhFlB...3.1746B}
}

@ARTICLE{Buchner_2006,
       author = {{B{\"u}chner}, J.},
        title = "{Theory and Simulation of Reconnection.  In memoriam Harry Petschek}",
      journal = {\ssr},
         year = 2006,
        month = jun,
       volume = {124},
       number = {1-4},
        pages = {345-360},
          doi = {10.1007/s11214-006-9094-x},
       adsurl = {https://ui.adsabs.harvard.edu/abs/2006SSRv..124..345B}
}

@ARTICLE{ChiouHau_2002,
       author = {{Chiou}, S. -W. and {Hau}, L. -N.},
        title = "{Tearing-mode instability in anisotropic plasmas: Effects of energy closure}",
      journal = {\grl},
         year = 2002,
        month = aug,
       volume = {29},
       number = {16},
          eid = {1815},
        pages = {1815},
          doi = {10.1029/2002GL014720},
       adsurl = {https://ui.adsabs.harvard.edu/abs/2002GeoRL..29.1815C}
}

@ARTICLE{ChiouHau_2003,
       author = {{Chiou}, S. -W. and {Hau}, L. -N.},
        title = "{Explosive and oscillatory tearing-mode instability in gyrotropic plasmas}",
      journal = {Physics of Plasmas},
         year = 2003,
        month = oct,
       volume = {10},
       number = {10},
        pages = {3813-3816},
          doi = {10.1063/1.1606682},
       adsurl = {https://ui.adsabs.harvard.edu/abs/2003PhPl...10.3813C}
}

@ARTICLE{Huang_2013hyper,
       author = {{Huang}, Yi-Min and {Bhattacharjee}, A. and {Forbes}, Terry G.},
        title = "{Magnetic reconnection mediated by hyper-resistive plasmoid instability}",
      journal = {Physics of Plasmas},
         year = 2013,
        month = aug,
       volume = {20},
       number = {8},
          eid = {082131},
        pages = {082131},
          doi = {10.1063/1.4819715},
archivePrefix = {arXiv},
       eprint = {1308.1871},
 primaryClass = {physics.plasm-ph},
       adsurl = {https://ui.adsabs.harvard.edu/abs/2013PhPl...20h2131H}
}

@ARTICLE{Shi_2020,
       author = {{Shi}, Chen and {Velli}, Marco and {Pucci}, Fulvia and {Tenerani}, Anna and {Innocenti}, Maria Elena},
        title = "{Oblique Tearing Mode Instability: Guide Field and Hall Effect}",
      journal = {\apj},
         year = 2020,
        month = oct,
       volume = {902},
       number = {2},
          eid = {142},
        pages = {142},
          doi = {10.3847/1538-4357/abb6fa},
archivePrefix = {arXiv},
       eprint = {2007.00607},
 primaryClass = {physics.plasm-ph},
       adsurl = {https://ui.adsabs.harvard.edu/abs/2020ApJ...902..142S}
}

@ARTICLE{Shay_1999,
       author = {{Shay}, M.~A. and {Drake}, J.~F. and {Rogers}, B.~N. and {Denton}, R.~E.},
        title = "{The scaling of collisionless, magnetic reconnection for large systems}",
      journal = {\grl},
         year = 1999,
        month = jul,
       volume = {26},
       number = {14},
        pages = {2163-2166},
          doi = {10.1029/1999GL900481},
       adsurl = {https://ui.adsabs.harvard.edu/abs/1999GeoRL..26.2163S}
}

@ARTICLE{Yamada_2010,
       author = {{Yamada}, Masaaki and {Kulsrud}, Russell and {Ji}, Hantao},
        title = "{Magnetic reconnection}",
      journal = {Reviews of Modern Physics},
         year = 2010,
        month = jan,
       volume = {82},
       number = {1},
        pages = {603-664},
          doi = {10.1103/RevModPhys.82.603},
       adsurl = {https://ui.adsabs.harvard.edu/abs/2010RvMP...82..603Y}
}

@ARTICLE{Mirnov_2004,
       author = {{Mirnov}, V.~V. and {Hegna}, C.~C. and {Prager}, S.~C.},
        title = "{Two-fluid tearing instability in force-free magnetic configuration}",
      journal = {Physics of Plasmas},
         year = 2004,
        month = sep,
       volume = {11},
       number = {9},
        pages = {4468-4482},
          doi = {10.1063/1.1773778},
       adsurl = {https://ui.adsabs.harvard.edu/abs/2004PhPl...11.4468M}
}

@ARTICLE{Hosseinpour_2009,
       author = {{Hosseinpour}, M. and {Bian}, N. and {Vekstein}, G.},
        title = "{Two-fluid regimes of the resistive and collisionless tearing instability}",
      journal = {Physics of Plasmas},
         year = 2009,
        month = jan,
       volume = {16},
       number = {1},
          eid = {012104},
        pages = {012104},
          doi = {10.1063/1.3068470},
       adsurl = {https://ui.adsabs.harvard.edu/abs/2009PhPl...16a2104H}
}

@ARTICLE{Meshcheriakov_2012,
       author = {{Meshcheriakov}, Dmytro and {Maget}, Patrick and {L{\"u}tjens}, Hinrich and {Beyer}, Peter and {Garbet}, Xavier},
        title = "{Linear stability of the tearing mode with two-fluid and curvature effects in tokamaks}",
      journal = {Physics of Plasmas},
         year = 2012,
        month = sep,
       volume = {19},
       number = {9},
          eid = {092509},
        pages = {092509},
          doi = {10.1063/1.4754000},
       adsurl = {https://ui.adsabs.harvard.edu/abs/2012PhPl...19i2509M}
}

@ARTICLE{Ugai_1992,
       author = {{Ugai}, M.},
        title = "{Computer studies on development of the fast reconnection mechanism for different resistivity models}",
      journal = {Physics of Fluids B},
         year = 1992,
        month = sep,
       volume = {4},
       number = {9},
        pages = {2953-2963},
          doi = {10.1063/1.860458},
       adsurl = {https://ui.adsabs.harvard.edu/abs/1992PhFlB...4.2953U}
}

@ARTICLE{Birn_2001,
       author = {{Birn}, Joachim and {Hesse}, Michael},
        title = "{Geospace Environment Modeling (GEM) magnetic reconnection challenge: Resistive tearing, anisotropic pressure and hall effects}",
      journal = {\jgr},
         year = 2001,
        month = mar,
       volume = {106},
       number = {A3},
        pages = {3737-3750},
          doi = {10.1029/1999JA001001},
       adsurl = {https://ui.adsabs.harvard.edu/abs/2001JGR...106.3737B}
}

@BOOK{Schindler2006,
       author = {{Schindler}, Karl},
        title = "{Physics of Space Plasma Activity}",
         year = 2006,
          doi = {10.2277/0521858976},
       adsurl = {https://ui.adsabs.harvard.edu/abs/2006pspa.book.....S}
}

@article{Park_1984_Prm,
    author = {Park, W. and Monticello, D. A. and White, R. B.},
    title = {Reconnection rates of magnetic fields including the effects of viscosity},
    journal = {The Physics of Fluids},
    volume = {27},
    number = {1},
    pages = {137-149},
    year = {1984},
    month = {01},
    issn = {0031-9171},
    doi = {10.1063/1.864502},
    url = {https://doi.org/10.1063/1.864502},
    eprint = {https://pubs.aip.org/aip/pfl/article-pdf/27/1/137/12743526/137_1_online.pdf},
}

@article{Ji_Daughton_2011,
    author = {Ji, Hantao and Daughton, William},
    title = {Phase diagram for magnetic reconnection in heliophysical, astrophysical, and laboratory plasmas},
    journal = {Physics of Plasmas},
    volume = {18},
    number = {11},
    pages = {111207},
    year = {2011},
    month = {10},
    issn = {1070-664X},
    doi = {10.1063/1.3647505},
    url = {https://doi.org/10.1063/1.3647505},
    eprint = {https://pubs.aip.org/aip/pop/article-pdf/doi/10.1063/1.3647505/14910995/111207_1_online.pdf},
}

@ARTICLE{KowalFalceta:2024,
       author = {{Kowal}, Grzegorz and {Falceta-Gon{\c{c}}alves}, Diego A.},
        title = "{Quenching of Tearing Mode Instability by Transverse Magnetic Fields in Reconnection Current Sheets}",
      journal = {arXiv e-prints},
         year = 2024,
        month = jul,
          eid = {arXiv:2407.09996},
        pages = {arXiv:2407.09996},
          doi = {10.48550/arXiv.2407.09996},
archivePrefix = {arXiv},
       eprint = {2407.09996},
 primaryClass = {astro-ph.SR},
       adsurl = {https://ui.adsabs.harvard.edu/abs/2024arXiv240709996K}
}

@ARTICLE{Derigs_etal:2018,
       author = {{Derigs}, Dominik and {Winters}, Andrew R. and {Gassner}, Gregor J. and {Walch}, Stefanie and {Bohm}, Marvin},
        title = "{Ideal GLM-MHD: About the entropy consistent nine-wave magnetic field divergence diminishing ideal magnetohydrodynamics equations}",
      journal = {Journal of Computational Physics},
         year = 2018,
        month = jul,
       volume = {364},
        pages = {420-467},
          doi = {10.1016/j.jcp.2018.03.002},
archivePrefix = {arXiv},
       eprint = {1711.06269},
 primaryClass = {physics.comp-ph},
       adsurl = {https://ui.adsabs.harvard.edu/abs/2018JCoPh.364..420D}
}
\bibliographystyle{aasjournal}

%% This command is needed to show the entire author+affiliation list when
%% the collaboration and author truncation commands are used.  It has to
%% go at the end of the manuscript.
%\allauthors

%% Include this line if you are using the \added, \replaced, \deleted
%% commands to see a summary list of all changes at the end of the article.
%\listofchanges

\end{document}